\documentclass[a4paper,11pt]{article}

\usepackage{jcappub}
\usepackage{amssymb}
\usepackage{graphicx}
\usepackage{amsmath}
\usepackage{hyperref}
\usepackage{subfigure}
\usepackage{multirow}
\usepackage{multicol}

\newcommand{\md}{\mathrm{d}}

\title{\boldmath Anisotropies of cosmological gravitational wave backgrounds in non-flat spacetime}

\author[a,b,e]{Rong-Gen Cai,}
\emailAdd{cairg@itp.ac.cn}

\author[b,d]{Shao-Jiang Wang,}
\emailAdd{schwang@itp.ac.cn}

\author[b,c]{Zi-Yan Yuwen}
\emailAdd{yuwenziyan@itp.ac.cn}

\author[b,c]{Xiang-Xi Zeng }
\emailAdd{zengxiangxi@itp.ac.cn}

\affiliation[a]{Institute of Fundamental Physics and 
Quantum Technology, Ningbo University, Ningbo, 315211, China}
\affiliation[b]{CAS Key Laboratory of Theoretical Physics, Institute of Theoretical Physics, Chinese Academy of Sciences (CAS), Beijing 100190, China}
\affiliation[c]{School of Physical Sciences, University of Chinese Academy of Sciences (UCAS), Beijing 100049, China}
\affiliation[d]{Asia Pacific Center for Theoretical Physics (APCTP), Pohang 37673, Korea}
\affiliation[e]{School of Fundamental Physics and Mathematical Sciences, Hangzhou Institute for Advanced Study (HIAS), University of Chinese Academy of Sciences (UCAS), Hangzhou 310024, China}

\abstract{Recent reports of stochastic gravitational wave background from four independent pulsar-timing-array collaborations have renewed the interest in the cosmological gravitational wave background (CGWB), which is expected to open a new window into the early Universe. Although the early Universe is supposed to be extremely flat from an inflationary point of view, the cosmic microwave background (CMB) data alone from the Planck satellite measurement prefers an enhanced lensing amplitude that can be explained by a closed Universe. In this paper, we propose an independent method to constrain the early-universe flatness from the anisotropies of CGWB. Using the generalized harmonic decompositions in the non-flat spacetime, we find CGWBs from different physical mechanisms such as cosmic inflation and phase transitions share the same integrated Sachs-Wolfe (ISW) term but possess different SW terms, which would exhibit different behaviors when including the spatial curvature since the ISW effect is more sensitive to the spatial curvature than the SW effect. Furthermore, we provide the cross-correlations between CGWB and CMB, implying a positive or negative correlation between their SW effect terms depending on the GW mechanisms, which may hint at the sign of $f_{\mathrm{NL}}$ when considering non-Gaussianity contributions to anisotropies.}

\begin{document}
\maketitle
\flushbottom

\section{Introduction}
\label{sec:intro}

A non-zero spatial curvature $K$ can play a crucial role in the evolution of our Universe, especially the large-scale physics and the topology of the Universe~\cite{Lachieze-Rey:1995qrb}. Recent Planck analysis of cosmic microwave background (CMB) anisotropies has pushed the standard cosmological model to an unprecedented precision~\cite{Planck:2018vyg} with $\Omega_{K} = -0.044^{+0.018}_{-0.015}$ (68\%, $Planck$ TT,TE,EE + lowE). Combining the Planck-CMB data with lensing and various BAO measurements can indeed constrain the spatial curvature to an extremely flat value $\Omega_{K} = 0.0007 \pm 0.0019$ (68\%, $Planck$ TT,TE,EE + lowE + lensing + BAO). However, these constraints rely on a specific cosmological model and arrive at an anomalously higher lensing amplitude~\cite{Handley:2019tkm,DiValentino:2019qzk,DiValentino:2020hov}. Nevertheless, a positive curvature can naturally explain the anomalously higher lensing amplitude and alleviate the tensions with the supernova observations at low redshifts~\cite{Handley:2019tkm,DiValentino:2019qzk,DiValentino:2020hov,Park:2018tgj} (see~\cite{Abdalla:2022yfr} for a review of tensions). Nevertheless, a non-flat Universe enhances the discordances with most of the late-time cosmological observables like BAO~\cite{DiValentino:2019qzk}. In addition to this, one cannot simply assume $\Omega_{K}=0$~\cite{Anselmi:2022uvj}. These motivate us to develop a new method to constrain the spatial curvature $\Omega_{K}$, and the anisotropies of cosmological gravitational wave backgrounds (CGWB) can be such a candidate.

As recent observations of the stochastic gravitational wave backgrounds (SGWB) by pulsar timing arrays (PTAs) from  NANOGrav~\cite{NANOGrav:2023gor}, EPTA~\cite{EPTA:2023fyk}, PPTA~\cite{Reardon:2023gzh}, and CPTA~\cite{Xu:2023wog} collaborations as well as the upcoming gravitational-wave (GW) detectors (such as the space-based detectors like LISA~\cite{LISA:2017pwj}, Taiji~\cite{Hu:2017mde,Ruan:2018tsw}, Tianqin~\cite{TianQin:2015yph}, and DECIGO~\cite{Kawamura:2006up}, and the ground-based detectors like Einstein Telescope~\cite{Maggiore:2019uih}, Cosmic Explorer~\cite{Reitze:2019iox}), we may be very close to the sensitivity of CGWBs (see~\cite{Caprini:2018mtu} for a review of CGWB). Similar to the CMB, the CGWB also exhibits anisotropies (see~\cite{LISACosmologyWorkingGroup:2022kbp} for a review). Due to the extremely weak interactions of gravitons with matters, the initial distribution function of gravitons is not thermal (unlike CMB) but depends on the production mechanisms such as inflation~\cite{Bartolo:2019oiq,Bartolo:2019yeu,Malhotra:2020ket,Adshead:2020bji,Dimastrogiovanni:2021mfs,Dimastrogiovanni:2019bfl}, phase transition~\cite{Geller:2018mwu,Kumar:2021ffi,Li:2021iva}, topological defects~\cite{Liu:2020mru,Cai:2021dgx,Jenkins:2018nty,Kuroyanagi:2016ugi}, and the productions of primordial black holes (PBHs)~\cite{Bartolo:2019zvb,Li:2023qua}. Therefore, different physical processes will leave different imprints on the anisotropies of CGWB, and in turn, the anisotropies of CGWB can be a very powerful tool to distinguish different production mechanisms. In addition, different spatial curvatures can also impact differently on the anisotropies of CGWB for different production mechanisms since different production mechanisms possess different SW effect terms while keeping the same ISW effect term, and the ISW effect term is more sensitive to the spatial curvature than the SW effect term. Therefore, it is necessary to conduct a complete survey on anisotropies of CGWB with spatial curvature, and in particular, for some representative production mechanisms like inflation, phase transition, and scalar-induced gravitational wave (SIGW).

There are extensive studies on CMB anisotropies in non-flat spacetime using the Boltzmann hierarchy equations~\cite{Hu:1997hp,Hu:1997mn,Tram:2013ima,Lesgourgues:2013bra,Pitrou:2020lhu}. The standard method is the so-called total angular momentum method~\cite{Hu:1997hp}. In Ref.~\cite{Hu:1997mn}, they generalize this method to arbitrary perturbation type and FLRW metric simply by replacing $\exp(i\vec{k}\cdot\vec{x})$ with an ansatz $\exp(i\delta(\vec{k},\vec{x}))$, which serves as a presumption in their paper (see Appendix.~\ref{app:TAM} for details). However, it is not obvious that we should consider a form like $\exp(i\delta(\vec{k},\vec{x}))$ with a unit norm, and we cannot simply derive the angular power spectrum $C_{l}^{\mathrm{GW}}$ without knowing the norm of $\exp(i\delta(\vec{k},\vec{x}))$. Therefore we propose a more direct method to obtain the final expression of $C_{l}^{\mathrm{GW}}$ with as few ansatzes as possible.

The paper is organized as follows. In Section~\ref{sec:line-of-sight method}, we generalize the line-of-sight integral method to a non-flat FLRW Universe. In Section~\ref{sec:correlations}, we first give rise to the angular power spectrum of SGWB and the initial conditions of different physical mechanisms, then we show our numerical results of the angular power spectra for different physical mechanisms and the cross-correlations between CGWB and CMB. Finally, the section~\ref{sec:condis} is devoted to conclusions and discussions. The Appendix~\ref{app:hierarchy} is given for the analysis of Boltzmann hierarchy equations, and the Appendix~\ref{app:TAM} is provided for the calculations using the total angular momentum method.

\section{Line-of-sight method in non-flat Universe}\label{sec:line-of-sight method}

In this section, we propose to improve the derivations for the angular power spectrum for CGWB from the line-of-sight integral of CGWB anisotropies in a non-flat spacetime with a constant curvature background, which is slightly different from but still equivalent to the standard total angular momentum method~\cite{Hu:1997hp,Hu:1997mn}. 

Under the short-wave approximation~\cite{Isaacson:1968a,Isaacson:1968b}, one could treat GWs as an ensemble of gravitons with a distribution function $f_{\mathrm{GW}}(\eta, \vec{x}, p, \hat{n})$, whose time evolution is controlled by the Boltzmann equation. Following Refs.~\cite{Bartolo:2019oiq,Bartolo:2019yeu}, the GW distribution function $f_\mathrm{GW}$ can be expanded on top of the isotropic background to the first order as
\begin{align}\label{eq:def_Gamma}
    f_\mathrm{GW}(\eta, \vec{x}, p, \hat{n}) = \bar{f}_\mathrm{GW}(p) - p \frac{\partial \bar{f}_\mathrm{GW}(p)}{\partial p} \Gamma(\eta, \vec{x}, p, \hat{n}),
\end{align}
where the minus sign in front of the anisotropy term follows the convention of the standard procedure of CMB. The background spacetime is described by a general FLRW metric with constant spatial curvature $K = -H_{0}^{2}(1-\Omega_{\mathrm{tot}})$,
\begin{align}
    \md s^2 &= a^2 \left( -\md \eta^2 + \frac{\md r^2}{1 - K r^2} + r^2\left(\md\theta^2 + \sin^2 \theta \md \phi^2\right) \right) \nonumber \\
    &= a^2 \left( -\md \eta^2 + \frac{1}{|K|}\left( \md \chi^2 + \sin_\mathrm{K}^2 \chi\left(\md\theta^2 + \sin^2 \theta \md \phi^2\right) \right)\right),
\end{align}
where $\sin_\mathrm{K} \chi\equiv\sqrt{|K|}r$ is the areal radius with introduction of the generalized sine function,
\begin{align}
    \sin_{\mathrm{K}} \chi = \left\{\begin{array}{ll}
        \sinh\chi, & K<0 \\
        \chi, & K\to 0 \\
        \sin\chi, & K>0
    \end{array}\right. ~.
\end{align}
Note here that for the case with $K\neq 0$, the traditional Fourier transformation is not well defined due to a non-vanishing length scale $|K|^{-1/2}$. W. Hu \textit{et al.}~\cite{Hu:1997mn} replaced the plane wave, $\exp(i\vec{k}\cdot\vec{x})$, in the Fourier transformation with a generalized one, $\exp(i\delta(\vec{k},\vec{x}))$, which is related to the eigenfunctions of Laplacian operators in the curved space, together with an ansatz on the recurrence equation of the multipole expansion of $\exp(i\delta(\vec{k},\vec{x}))$. However, it is not proved that a generalized plane wave satisfying the ansatz does possess a unit norm, i.e. $|\exp(i\delta(\vec{k},\vec{x}))|=1$. This motivates us to find a better method to perform the generalized Fourier transformation with as fewer assumptions as possible.

\subsection{Generalized Fourier transformation}

In this subsection, we provide a detailed description of a generalized Fourier transformation on a $3$-dimensional Riemann manifold $M^{(3)}_K$ with a constant curvature $K$ in the polar coordinate $\{\chi, \theta, \phi\}$ with the line element of the form,
\begin{align}
    \md l^2 = \gamma_{ij}\md x^i\md x^j =  \md \chi^2 + \sin_\mathrm{K}^2 \chi \left(\md\theta^2 + \sin^2 \theta \md \phi^2\right).
\end{align}
Consider a unit vector $\hat{n}$ in the tangent space $TM^{(3)}_K|_{\vec{x}}$, any $f$ as a scalar function of $(\vec{x}, \hat{n})$ can be expanded in the generalized harmonic modes with respect to $\vec{x}$,
\begin{align}\label{eq:H_modes_expansion}
    f(\vec{x}, \hat{n}) = \frac{1}{(2\pi)^3} \int \md m(\beta) \sum_{l=0}^{l_\mathrm{max}} \sum_{m=-l}^l \tilde{f}^{(1)}_{lm}(\beta,\hat{n}) \Phi_{\beta}^{~l} (\chi) Y_{lm}(\hat{x}),
\end{align}
where $\Phi_{\beta}^{~l}$ is the hyper-spherical Bessel function~\cite{Lifshitz:1963,Abbott:1986ct,Hu:1997mn,Tram:2013ima}. Together with the spherical harmonics, $\Phi_{\beta}^{~l} (\chi) Y_{lm}(\hat{x})$ serves as an eigenfunction for the Laplace operator on $M_K^{(3)}$,
\begin{align}
    \left( D_i D^i + k^2 \right) \Phi_{\beta}^{~l} (\chi) Y_{lm}(\hat{x}) = 0.
\end{align}
Note that the eigenfunctions are labeled with an index $\beta$ rather than the comoving wave-number $k$. For $K=0$, they are trivially given by $\beta = k$ and $\Phi_{\beta}^{~l} (\chi) = j_l(kr)$. For $K\neq 0$, it holds $|K|\beta^2 = k^2 + K$ with $\beta\geq 0$. Specifically, $\beta$ only takes integer values for $K>0$ due to the periodic boundary condition on $\chi$. Furthermore, it has been shown in Ref.~\cite{Lifshitz:1963} that the cases with $\beta=1$, $2$ correspond to some pure gauge modes so that the spectrum for the eigenvalues starts with an integer greater than or equal to three, $3\leq\beta\in\mathbb{Z}$. The wave-number space measure $m(\beta)$~\cite{Abbott:1986ct} is defined in Table.~\ref{tab:measure}. The upper bound for the summation over $l$ is given by $l_\mathrm{max} = \beta - 1$ for $K>0$ (closed geometry) and $l_\mathrm{max} = \infty$ for $K\leq 0$ (flat and open geometries).

\begin{table}[!tbp]
    \caption{The measure in wave-number space}
    \label{tab:measure}
    \centering
    \setlength{\tabcolsep}{6pt}% column separation
    \renewcommand{\arraystretch}{3}%row space 
    \begin{tabular}{cccc}
        \hline
        Curvature & $\displaystyle\int\md m(\beta)$ & \multicolumn{2}{c}{Delta function $\displaystyle \frac{1}{\beta^2}\delta(\beta,\beta')$}\\
        \hline
        $K<0$ & $\displaystyle\int_0^\infty\beta^2\md\beta$ & \quad $\displaystyle\frac{1}{\beta^2}\delta(\beta - \beta')$ & Radial Dirac Delta \\
        % \hline
        $K=0$ & $\displaystyle\int_0^\infty\beta^2\md\beta$ & \quad $\displaystyle\frac{1}{\beta^2}\delta(\beta - \beta')$ & Radial Dirac Delta \\
        $K>0$ & $\displaystyle\sum_{\beta=3}^\infty\beta^2$ & \quad $\displaystyle\frac{1}{\beta^2}\delta_{\beta, \beta'}$ & Radial Kronecker Delta \\
        \hline
    \end{tabular}
\end{table}

The coefficient $\tilde{f}^{(1)}_{lm}(\beta,\hat{n})$ can be expressed in terms of a harmonic inverse transformation,
\begin{align}\label{eq:re_FT_coefficient}
    \tilde{f}^{(1)}_{lm}(\beta,\hat{n}) = \int\md^2\hat{k}~ \tilde{f}^{(2)}(\beta,\hat{n},\hat{k}) Y_{lm}^*(\hat{k}) = 
    \int\md^2\hat{k}~ \tilde{f}^{(3)}(\beta,\hat{n},\hat{k})\cdot 4\pi i^l Y_{lm}^*(\hat{k}),
\end{align}
where $\tilde{f}^{(2)}$ and $\tilde{f}^{(3)}$ is simply related by $\tilde{f}^{(2)}=4\pi i^l\tilde{f}^{(3)}$, and $\hat{k}$ is the unit tangent vector in $TM^{(3)}_K|_{\vec{x}}$. Since both $\hat{n}$ and $\hat{k}$ are two unit vectors in the same vector space, the relative angle between them is well defined. The cosine of the relative angle is denoted as $\mu=\hat{k}\cdot\hat{n}$. Since the angular expansion \eqref{eq:re_FT_coefficient} is symmetric along the direction of $\hat{n}$, different directions with the same $\mu$ are equivalent, thus the expansion coefficients can be regarded as functions of $\hat{k}$ and the relative orientation, $\tilde{f}^{(3)}(\beta,\hat{n},\hat{k}) \equiv \tilde{f}(\beta,\mu,\hat{k})$. Plugging \eqref{eq:re_FT_coefficient} into \eqref{eq:H_modes_expansion} gives rise to
\begin{align}
    f( \vec{x}, \hat{n}) = \frac{1}{(2\pi)^3} \int \md m(\beta) \int \md^2 \hat{k} ~ \tilde{f}(\beta,\mu,\hat{k}) \sum_{l,m} 4\pi i^l    Y_{lm}^*(\hat{k}) \Phi_{\beta}^{~l}(\chi) Y_{lm}(\hat{x}).
\end{align}
Finally, by defining the ``mode'' function labeled by $\beta$ and $\hat{k}$ as
\begin{align}
    M_\beta(\hat{k},\chi,\hat{x}) =  4\pi \sum_{l=0}^{l_{\mathrm{max}}} \sum_{m=-l}^l i^l ~Y_{lm}^*(\hat{k}) \Phi_{\beta}^{~l}(\chi) Y_{lm}(\hat{x}).
\end{align}
we can write down a generalized Fourier transformation of scalar function $f$ in terms of the mode functions $M_\beta$ as
\begin{align} \label{eq:G_Fourier_Transform}
    f( \vec{x}, \hat{n}) = \frac{1}{(2\pi)^3} \int \md m(\beta) \int \md^2 \hat{k} ~ \tilde{f}(\beta,\mu,\hat{k}) M_\beta(\hat{k},\chi,\hat{x}).
\end{align}
It is easy to check that such an expansion returns to the usual Fourier transformation for $K=0$,
\begin{align}
    f( \vec{x}, \hat{n}) &= \frac{1}{(2\pi)^3} \int_0^\infty k^2\md k \int \md^2 \hat{k} ~ \tilde{f}(k,\mu,\hat{k}) M_\beta(k, \hat{k}, r, \hat{x}) = \int \frac{\md^3 \vec{k}}{(2\pi)^3} ~ \tilde{f}(\mu,\vec{k}) e^{i\vec{k}\cdot\vec{x}},
\end{align}
where we have degenerate the mode functions back to the usual Fourier modes,
\begin{align}
    e^{i\vec{k}\cdot\vec{x}} = 4\pi \sum_{l=0}^{\infty} \sum_{m=-l}^l i^l ~Y_{lm}^*(\hat{k}) j_l(kr) Y_{lm}(\hat{x}).
\end{align}

\begin{figure}
    \centering
    \includegraphics[width = 0.55\textwidth]{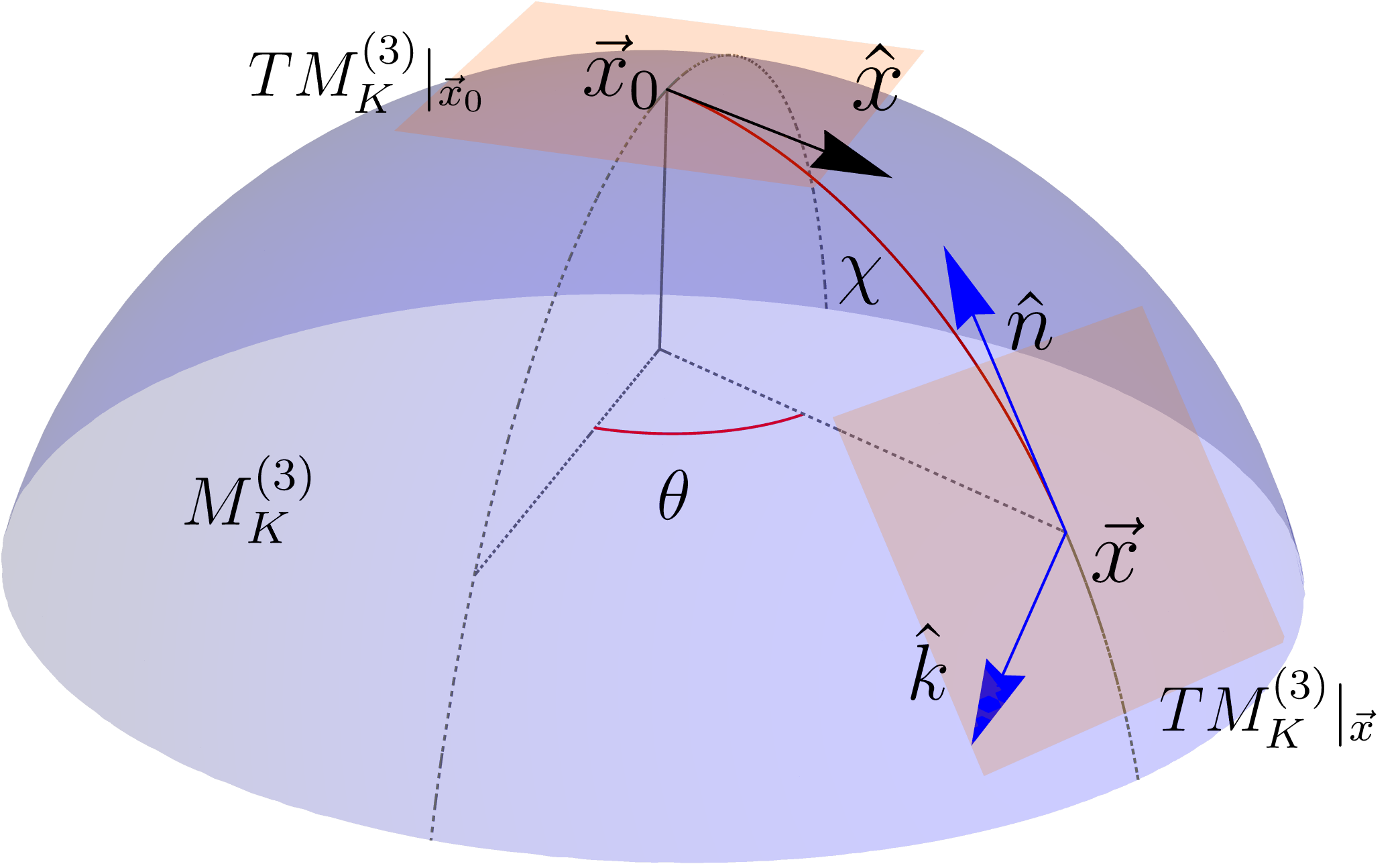}
    \quad 
    \includegraphics[width = 0.35\textwidth]{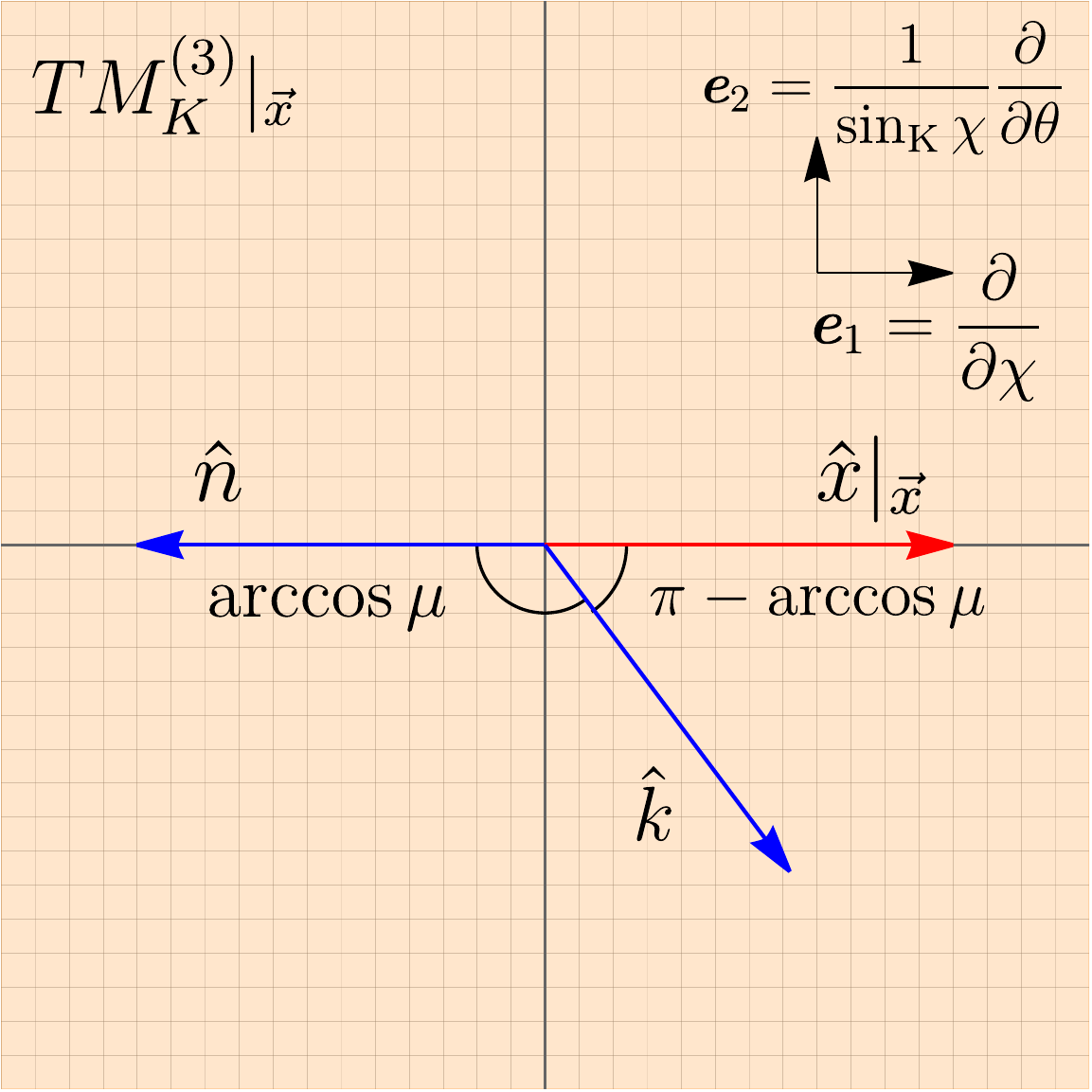}
    \caption{An intuitive illustration of a $2$-dimensional slice embedded in a flat space with one higher dimension (left) and the corresponding tangent space of $\vec{x}$ (right), where the spatial geometry is set to be closed ($K>0$) as an example. Such a picture can be interpreted as the $3$-dimensional manifold $M^{(3)}_K$ with the $\phi$ coordinate compressed.}
    \label{fig:spatial_slice}
\end{figure}

The direction $\hat{x}$ has a two-fold meaning: it is not only the angular coordinate $(\theta_x, \phi_x)$ of $\vec{x}$, but also a unit vector living in the tangent space of the center point as shown in Fig.~\ref{fig:spatial_slice} from a $2$-dimensional point of view. In our formalism, a specific coordinate system $\{\chi, \theta, \phi\}$ has been chosen with the observer staying at the origin $\vec{x}_0$ namely $\chi = 0$. Similar to CMB, GWs from $\vec{x}$ that can be detected by an observer at $\vec{x}_0$ must travel towards the direction $\hat{n}$ and along the $\chi$-coordinate line. One could parallel transport $\hat{n}$ to $\vec{x}_0$ along the geodesic and get $\hat{n}|_{\vec{x}_0} = - \hat{x}$, which should remain tangent to the $\chi$-coordinate line along the whole path. Alternatively, one could also transport the tangent vectors in $TM^{3}_K|_{\vec{x}_0}$ to $TM^{3}_K|_{\vec{x}}$, which leads to an equivalent result $\hat{x}|_{\vec{x}} = - \hat{n}$. Therefore, the directional derivative on a scalar function $F$ along $\hat{n}$ can be rewritten as partial derivative,
\begin{align}\label{eq:ni_Di}
    \hat{n}^i D_i F = \hat{n}^i \partial_i F = -\frac{\partial F}{\partial \chi},
\end{align}
Note that the conclusions above remain valid even after performing a conformal transformation $\md s^2_{(3)} = \frac{a(t)^2}{|K|} \md l^2$ with $a(t)$ and $K$ independent of the spatial coordinates, since the conformal transformation only rescales the metric but merely changes the structure of tangent space and radial geodesics. In this case, the directional derivative is modified as
\begin{align}\label{eq:ni_Di_modified}
    \hat{n}^i D_i F = \hat{n}^i \partial_i F = - \frac{\sqrt{|K|}}{a} \frac{\partial F}{\partial \chi},
\end{align}
which will be used shortly below in the next subsection when solving the Boltzmann equation.

\subsection{Line-of-sight integral}

The evolution of the distribution function of GWs can be solved from the Boltzmann equation,
\begin{align}
    \frac{\md}{\md \eta}f_\mathrm{GW}(x^\mu, p^\mu) = \mathcal{C}\left( f_\mathrm{GW}(x^\mu, p^\mu) \right) + \mathcal{T}\left( f_\mathrm{GW}(x^\mu, p^\mu) \right),
\end{align}
where $\mathcal{C}$ and $\mathcal{T}$ denote the scattering term and emission term, respectively. However, unlike CMB, the scattering effect is negligible due to the extremely weak couplings between GWs and other components of the cosmic fluids. Meanwhile, for the GWs generated deep into the very early Universe, the emission term only matters at the initial time, which can be treated as an initial condition and dropped away when solving the time evolution. Therefore, the Boltzmann equation takes a simple form $\md f_\mathrm{GW} / \md \eta = 0$, whose solution is obtained as follows.

\subsubsection{Boltzmann equation to the leading-order in scalar perturbations} 

Considering only the scalar perturbations $\Psi(\eta,\vec{x})$ and $\Phi(\eta,\vec{x})$ on the FLRW background, the line element reads
\begin{align}
    \md s^2 &= a^2 \left( -(1+2\Psi)\md \eta^2 + \frac{1-2\Phi}{|K|}\gamma_{ij}\md x^i\md x^j\right) \nonumber\\
    &= a^2 \left( -(1+2\Psi)\md \eta^2 + \frac{1-2\Phi}{|K|}\left( \md \chi^2 + \sin_\mathrm{K}^2 \chi\left(\md\theta^2 + \sin^2 \theta \md \phi^2\right) \right)\right),
\end{align}
under which the Boltzmann equation becomes~\cite{Bartolo:2019oiq,Bartolo:2019yeu}
\begin{align}
    \frac{\md f_\mathrm{GW}}{\md \eta} = \frac{\partial f_\mathrm{GW}}{\partial \eta} + \frac{\md x^i}{\md \eta} \partial_i f_\mathrm{GW} + p \frac{\partial f_\mathrm{GW}}{\partial p}\left( \frac{\partial \Phi}{\partial \eta} - \frac{\md x^i}{\md \eta} \partial_i \Psi \right) = 0.
\end{align}
Using the definition~\eqref{eq:def_Gamma}, the Boltzmann equation to the leading order becomes
\begin{align}\label{eq:Boltzmann_for_Gamma}
    \frac{\partial \Gamma}{\partial \eta} + \frac{\md x^i}{\md \eta} \partial_i \Gamma = \frac{\partial \Phi}{\partial \eta} - \frac{\md x^i}{\md \eta} \partial_i \Psi.
\end{align}

\subsubsection{Perturbed Boltzmann equation after generalized Fourier transformation} 

We then perform the generalized Fourier transformation~\eqref{eq:G_Fourier_Transform} to extract each mode with its time dependence in the expansion coefficients and the space dependence in the mode functions, then the resulting equation that each mode should satisfy now becomes
\begin{align}\label{eq:Boltzmann_for_Gamma_tilde}
    \frac{\partial \tilde{\Gamma}}{\partial \eta}M_\beta + a \hat{n}^i\tilde{\Gamma} \partial_i M_\beta =  \frac{\partial \tilde{\Phi}}{\partial \eta} M_\beta - a \hat{n}^i\tilde{\Psi} \partial_i M_\beta.
\end{align}
Here we have used the fact that $\hat{n}^i \propto \md x^i / \md \eta$, where the proportion factor is determined by the normalization of unit vector $\frac{a^2}{|K|}\gamma_{ij} \hat{n}^i \hat{n}^j = 1$ and the leading-order null geodesic equation $\md\eta^2 = \gamma_{ij}\md x^i\md x^j / |K|$. 

It is intriguing to note that, the directional derivative operator on both sides of the Boltzmann equation~\eqref{eq:Boltzmann_for_Gamma_tilde} can be further converted into a time derivative. To see this, note that for GWs freely traveling along a radial null geodesic, the world line to the leading order can be parameterized as $\chi = \sqrt{|K|}(\eta_0 - \eta)$ with respect to the conformal time today $\eta_0$, then $\partial_\chi = -|K|^{-1/2} \partial_\eta$ together with Eq.~\eqref{eq:ni_Di_modified} can rewrite the directional derivative operator as
\begin{align}
    a \hat{n}^i\partial_i = a \frac{-\sqrt{|K|}}{a} \frac{\partial }{\partial\chi} =  -\sqrt{|K|}\frac{\partial \eta}{\partial\chi} \frac{\partial}{\partial\eta} = \frac{\partial}{\partial\eta},
\end{align}
with which the Boltzmann equation can be further rearranged into total derivative terms as
\begin{align}
    \frac{\partial}{\partial\eta} &
    \left(\tilde{\Gamma}(\eta,\beta,\mu,\hat{k}) M_\beta(\hat{k},\chi,\hat{x})
    \right) \nonumber\\
    &= -\frac{\partial}{\partial\eta}\left(\tilde{\Psi} (\eta,\beta,\hat{k}) M_\beta(\hat{k},\chi, \hat{x})\right) + 
    \left( \frac{\partial \tilde{\Phi}(\eta,\beta,\hat{k})}{\partial \eta} + 
    \frac{\partial \tilde{\Psi}(\eta,\beta,\hat{k})}{\partial \eta} \right)
    M_\beta(\hat{k},\chi,\hat{x}).
\end{align}

Note here that the scalar perturbations $\tilde{\Psi}$ and $\tilde{\Phi}$ do not contain the $\mu$ dependence, thus only perturbations that are equivalent in each direction are considered. Note also that the dependence on momentum $p$ does not enter in the following discussions, hence we temporarily leave out the explicit $p$ dependence in this subsection but recover it in the next section.

Now, we can directly integrate over conformal time from initial time $\eta_\mathrm{in}$ to nowadays $\eta_0$ to solve for 
\begin{align}\label{eq:line-of-sight}
    \tilde{\Gamma}(\eta_0,\beta,\mu,\hat{k}) M_\beta(\hat{k},\chi_0,\hat{x}) =& \left(
    \tilde{\Gamma}(\eta_\mathrm{in},\beta,\mu,\hat{k})
    + \tilde{\Psi}(\eta_\mathrm{in},\beta,\hat{k})
    \right)M_\beta(\hat{k},\chi_\mathrm{in},\hat{x}) \nonumber\\
    &\quad + \int_{\eta_\mathrm{in}}^{\eta_0}\md\eta 
    \left( 
    \frac{\partial \tilde{\Phi}(\eta,\beta,\hat{k})}{\partial \eta} + 
    \frac{\partial \tilde{\Psi}(\eta,\beta,\hat{k})}{\partial \eta}
    \right) M_\beta(\hat{k},\chi,\hat{x}),
\end{align}
where we have dropped $\tilde{\Psi}$ at boundary $\eta_0$. This is the so-called line-of-sight integration approach \cite{Contaldi:2016koz,Seljak:1996is}. Using the property of hyper-spherical Bessel function $\Phi_\beta^{~l}(0) = \delta_{l,0}$, the mode function at $\eta_0$ is given by
\begin{align}
    M_\beta(\hat{k},\chi_0,\hat{x}) &= 4\pi \sum_{l,m}i^l Y_{lm}^*(\hat{k})
    \Phi_\beta^{~l}\left(\sqrt{|K|}(\eta_0-\eta_0)\right)
    Y_{lm}(\hat{x}) \nonumber\\
    &= 4\pi \sum_{l,m}i^l Y_{lm}^*(\hat{k})Y_{lm}(\hat{x}) \delta_{l,0} = 4\pi Y_{00}^*(\hat{k})Y_{00}(\hat{x}) = 1.
\end{align}
Thus, the left-hand side of Eq.~\eqref{eq:line-of-sight} reduces to $\tilde{\Gamma}|_{\eta_0}$.

\subsubsection{Fourier-transformed Boltzmann equation after projection}

By multiplying both sides with $\frac{i^l}{2}P_l(\mu)$ and integrating $\mu$ from $-1$ to $1$, the Boltzmann equation~\eqref{eq:line-of-sight} can be projected to multipole moments labeled by $l$, with the projected anisotropy $\tilde{\Gamma}_{l}$ defined as
\begin{align}
    \tilde{\Gamma}_{l}(\eta,\beta,\hat{k}) &\equiv \frac{i^l}{2}\int_{-1}^1 \md\mu ~ P_l(\mu) \tilde{\Gamma}(\eta,\beta,\mu,\hat{k}), \\
    \tilde{\Gamma}(\eta,\beta,\mu,\hat{k}) &= \sum_{l}(-i)^l(2l+1)P_l(\mu) \tilde{\Gamma}_{l}(\eta,\beta,\hat{k}),
\end{align}
where $P_l(\mu)$ is the Legendre polynomials and $\mu = \hat{k}\cdot \hat{n}$.
Similarly, the projection of mode function $M_\beta$ is given by
\begin{align}
    \mathrm{Proj}_l M_\beta(\hat{k},\chi,\hat{x}) &\equiv \frac{i^l}{2}\int_{-1}^1 \md\mu ~ P_l(\mu) M_\beta(\hat{k},\chi,\hat{x}) \nonumber \\
    & = \frac{i^l}{2}\int_{-1}^1 \md\mu ~ P_l(\mu) 
    4\pi \sum_{{l'},m}i^{l'} Y_{{l'}m}^*(\hat{k})
    \Phi_\beta^{~{l'}}\left(\sqrt{|K|}(\eta_0-\eta)\right)
    Y_{{l'}m}(-\hat{n}).
\end{align}
Since $M_\beta(\hat{k},\chi,\hat{x})$ is a well-defined scalar function on $M^{(3)}_K$, and it should be also independent of the choice of coordinate, then the dependence on $\hat{k}$ and $\hat{x} = -\hat{n}$ can be transferred to the dependence on $\mu$. One can thus choose a specific coordinate such that $\hat{n}$ points along $\theta = 0$ and $\phi = 0$ to simplify our calculation, in which case the spherical harmonics are given by
\begin{align} \label{eq:Ylm_khat}
    Y_{lm}^*(\hat{k}) &= \sqrt{\frac{2l+1}{4\pi} \frac{(l-m)!}{(l+m)!}} P_l^m(\mu) e^{-im\phi_{\hat{k}}}, \\
    Y_{lm}(\hat{x}) &= \sqrt{\frac{2l+1}{4\pi} \frac{(l-m)!}{(l+m)!}} P_l^m(-1) e^{im\phi_{\hat{x}}} =
    \sqrt{\frac{2l+1}{4\pi}} (-1)^l \delta_{m,0}.
    \label{eq:Ylm_xhat}
\end{align}
In this coordinate the projection reads
\begin{align}
    \mathrm{Proj}_l M_\beta(\hat{k},\chi,\hat{x}) &= 4\pi\sum_{{l'}m} \Phi_\beta^{~{l'}} \frac{i^{l+l'}}{2} \frac{2l+1}{4\pi} e^{-im\phi_{\hat{k}}} (-1)^{l'}\delta_{m,0} \int_{-1}^1\md\mu P_l(\mu) P_{l'}^m(\mu) \nonumber \\
    & = \sum_{l'} \Phi_\beta^{~{l'}} \frac{2l+1}{2} i^{l+l'} (-1)^l \cdot \frac{2}{2l'+1} \delta_{l,l'} = \Phi_\beta^{~l}.
\end{align}
It is easy to see that, by choosing the coordinate axis fixed by~\eqref{eq:Ylm_khat} and~\eqref{eq:Ylm_xhat}, the mode function $M_\beta$ can be expressed in the following form,
\begin{align}\label{eq:M_beta_of_mu}
    M_\beta(\hat{k},\chi,\hat{x}) = \sum_{l} (-i)^l (2l+1) \Phi_\beta^{~l}(\chi) P_l(\mu).
\end{align}
Finally, the projection of line-of-sight integral \eqref{eq:line-of-sight} can be expressed as
\begin{align}\label{eq:Gamma_l}
    \tilde{\Gamma}_{l}(\eta,\beta,\hat{k}) =& 
    \frac{i^l}{2} \int_{-1}^{1}\md\mu ~ P_l(\mu) \tilde{\Gamma} (\eta_\mathrm{in},\beta,\mu,\hat{k}) M_\beta(\hat{k},\chi_\mathrm{in},\hat{x})
    + \tilde{\Psi}(\eta_\mathrm{in},\beta,\hat{k})
    \Phi_\beta^{~l}\left(\sqrt{|K|}(\eta_0-\eta_\mathrm{in})\right) \nonumber\\
    & \quad + \int_{\eta_\mathrm{in}}^{\eta_0}\md\eta 
    \left( 
    \frac{\partial \tilde{\Phi}(\eta,\beta,\hat{k})}{\partial \eta} + 
    \frac{\partial \tilde{\Psi}(\eta,\beta,\hat{k})}{\partial \eta}
    \right) \Phi_\beta^{~l}\left(\sqrt{|K|}(\eta_0-\eta)\right),
\end{align}
where the first term on the right-hand side contains all multipole contributions at initial time, and the last two terms represent the gravitational effects.

\subsubsection{Hierarchy Boltzmann equations after multipole decomposition}

Here we first discuss more about the initial multipole contribution by writing down the initial multipole terms on the right-hand side of Eq.~\eqref{eq:line-of-sight} explicitly,
\begin{align}\label{eq:Gamma*M}
    \tilde{\Gamma} (\eta_{\mathrm{in}},\beta,\mu,\hat{k}) &M_\beta(\hat{k},\chi_{\mathrm{in}},\hat{x}) = \sum_{l_1} (-i)^{l_1}(2l_1+1)P_{l_1}(\mu) \tilde{\Gamma}_{l_1} \sum_{l_2, m} 4\pi i^{l_2} Y_{l_2 m}^*(\hat{k}) \Phi_\beta^{~l_2} Y_{l_2 m}(\hat{x}) \nonumber\\
    & = \sum_{l_1,l_2}(-i)^{l_1 + l_2} (2l_1+1)(2l_2+1) P_{l_1}(\mu) P_{l_2}(\mu) \tilde{\Gamma}_{l_1} \Phi_{\beta}^{~l_2}.
\end{align}
A projection can be implemented for Eq.~\eqref{eq:Gamma*M} by multiplying $\frac{i^l}{2}P_l(\mu)$ and performing an integral over $\mu$,
\begin{align}
    \frac{i^l}{2} & \int_{-1}^{1}\md\mu ~ P_l(\mu) \tilde{\Gamma} (\eta_\mathrm{in},\beta,\mu,\hat{k}) M_\beta(\hat{k},\chi_\mathrm{in},\hat{x}) \nonumber \\
    &= \sum_{l_1,l_2}\tilde{\Gamma}_{l_1} \Phi_{\beta}^{~l_2} (-i)^{l_1 + l_2} (2l_1+1)(2l_2+1) \frac{i^l}{2}\int_{-1}^{1}\md\mu ~ P_l(\mu) P_{l_1}(\mu) P_{l_2}(\mu) \nonumber\\
    & = \sum_{l_1,l_2}\tilde{\Gamma}_{l_1} \Phi_{\beta}^{~l_2} (-i)^{l_1 + l_2} i^l \frac{(2l_1+1)(2l_2+1)(4\pi)^{3/2}}{\sqrt{(2l_1+1)(2l_2+1)(2l+1)}} \frac{1}{4\pi}\int\md^2\hat{n} Y_{l_10}(\hat{n})Y_{l_20}(\hat{n})Y_{l0}(\hat{n}) \nonumber \\
    & = \sum_{l_1,l_2}\tilde{\Gamma}_{l_1} \Phi_{\beta}^{~l_2} (-1)^{(l_1+l_2 - l)/2}
    (2l_1+1)(2l_2+1) 
    \left(\begin{array}{ccc}
        l_1 & l_2 & l \\
        0 & 0 & 0
    \end{array}\right) 
    \left(\begin{array}{ccc}
        l_1 & l_2 & l \\
        0 & 0 & 0
    \end{array}\right)~,
    \label{eq:Gamma*M_projected}
\end{align}
where the Wigner $3$j-symbols are related to the Clebsch-Gordan coefficients by
\begin{align}
    \left(\begin{array}{ccc}
        l_1 & l_2 & l \\
        m_1 & m_2 & m
    \end{array}\right) = \frac{(-1)^{l_1 - l_2 - m}}{\sqrt{2L+1}} \left[\begin{array}{ccc}
        l_1 & l_2 & l \\
        m_1 & m_2 & -m
    \end{array}\right] ~.
\end{align}
In particular, for $m_1=m_2=m=0$, the non-vanishing condition requires $l_1 + l_2 + l$ is even, or equivalently $l_1 + l_2 - L$ is even, which ensures the factor $(-1)^{(l_1+l_2 - l)/2}$ in the last line of~\eqref{eq:Gamma*M_projected} to remain real. It is easy to see from Eq.~\eqref{eq:Gamma*M_projected} that the multipole  mode $\tilde{\Gamma}_l |_{\eta_0}$ observed today is affected by all multipoles at initial time $\tilde{\Gamma}_{l_1} |_{\eta_\mathrm{in}}$ via a summation over $l_1$. 

Considering a fixed $l_1$ in the summation corresponding to a specific multipole-mode labeled by $l_1$, the coefficients are determined by hyper-spherical Bessel functions labeled by $l_2$ and Clebsch-Gordan coefficients, which adds an additional constraint on $l_1$, $l_2$ and $l$ by the selection rules. For example, the monopole term $l_1 = 0$ reads
\begin{align}
    \sum_{l_2}\tilde{\Gamma}_{0} \Phi_{\beta}^{~l_2} (-1)^{(l_2 - l)/2}
    (2l_2+1) 
    \left(\begin{array}{ccc}
        0 & l_2 & l \\
        0 & 0 & 0
    \end{array}\right)^2
    % \left(\begin{array}{ccc}
    %     0 & l_2 & l \\
    %     0 & 0 & 0
    % \end{array}\right)
    = \tilde{\Gamma}_{0} \Phi_{\beta}^{~l} ~,
\end{align}
which is just the first term in Eq.~\eqref{eq:Gamma_l}. The dipole term $l_1 = 1$ reads
\begin{align}
    \sum_{l_2} &~ \tilde{\Gamma}_{1} \Phi_{\beta}^{~l_2} (-1)^{(1+ l_2 - l)/2}
    3(2l_2+1) 
    \left(\begin{array}{ccc}
        1 & l_2 & l \\
        0 & 0 & 0
    \end{array}\right)^2
    % \left(\begin{array}{ccc}
    %     1 & l_2 & l \\
    %     0 & 0 & 0
    % \end{array}\right) 
    \nonumber\\
    &= 3\tilde{\Gamma}_{1} \Phi_\beta^{~l_2} 
    \left(
        \delta_{l_2, l-1} \frac{l_2+1}{2l_2 +3} - 
        \delta_{l_2, l+1} \frac{l_2}{2l_2 - 1}
    \right) = \frac{3\tilde{\Gamma}_{1}}{2l+1} \left( l \Phi_\beta^{~l-1} - (l+1) \Phi_\beta^{~l+1} \right)  \nonumber \\
    & = 3\tilde{\Gamma}_1 \left(
        \left(\frac{l+1}{2l+1} \frac{\sqrt{\beta^2 - \hat{K} l^2}}{\sqrt{\beta^2 - \hat{K}(l+1)^2}} + \frac{l}{2l+1}\right) \Phi_\beta^{~l-1} - \frac{(l+1)\cot_{\mathrm{K}}\chi}{\sqrt{\beta^2 - \hat{K}(l+1)^2}} \Phi_\beta^{~l}
    \right),
\end{align}
where in the last line we have used the recurrence relation,
\begin{align}
    \sqrt{\beta^2 - \hat{K}l^2} \Phi_\beta^{~l}(\chi) &= (2l-1) \cot_\mathrm{K}\chi~\Phi_\beta^{~l-1}(\chi) -  \sqrt{\beta^2 - \hat{K}(l-1)^2} ~\Phi_\beta^{~l-2}(\chi),
\end{align}
with
\begin{align}
    \cot_\mathrm{K}\chi &\equiv \left\{\begin{array}{ll}
        \coth\chi, & K<0 \\
        1/\chi, & K\to 0 \\
        \cot\chi, & K>0
    \end{array}\right. ~.
\end{align}
This result shares the same formula as the Doppler effect in CMB since they both account for the initial dipole perturbation. Here the Doppler effect in CMB is induced by the Thompson scattering between photons and baryons, which carries the velocity perturbation information of baryons. Since there is no significant scattering between GWs and other components of the cosmic fluid, the Doppler effect directly implies the velocity perturbation of the GW sources. Following the same procedure, every multipole contribution can be in principle evaluated. 

Back to the Boltzmann equation \eqref{eq:Boltzmann_for_Gamma}, performing multipole decomposition provides the hierarchy equation as
\begin{align}\label{eq:hierarchy_eq}
    \frac{\partial \tilde{\Gamma}_{l}}{\partial\eta} 
    - \frac{l s_{l}}{2l+1} \tilde{\Gamma}_{l-1} 
    + \frac{(l+1) s_{l+1}}{2l+1} \tilde{\Gamma}_{l+1} =  \frac{\partial \tilde{\Phi}}{\partial\eta} \delta_{l,0} + \frac{s_1}{3} \tilde{\Psi} \delta_{l,1} 
\end{align}
with an abbreviation $s_{l} \equiv \sqrt{|K|} \sqrt{\beta^2 - \hat{K}l^2}$ for $K\neq 0$ and $s_l\equiv \beta$ for $K=0$. The detailed derivation is presented in Appendix.~\ref{app:hierarchy}. Here Eq.~\eqref{eq:hierarchy_eq} suggests an hierarchy of multipole contributions $\tilde{\Gamma}_l\sim k\eta \tilde{\Gamma}_{l-1}$. Since all the modes that we are interested in are on super-horizon scales $k\eta_\mathrm{in}\ll 1$, higher modes with $l\geq 1$ can be safely dropped away. To the next-to-leading order, Eq.~\eqref{eq:Gamma_l} can be rewritten as
\begin{align}\label{eq:Gammal_SW_and_ISW}
    \tilde{\Gamma}_{l}(\eta_0,\beta,\hat{k}) \simeq& \underbrace{ \left(\tilde{\Gamma}_0(\eta_\mathrm{in},\beta,\hat{k}) + \tilde{\Psi}(\eta_\mathrm{in},\beta,\hat{k}) \right)}_{\mathrm{SW}} \Phi_\beta^{~l}\left(\sqrt{|K|}(\eta_0-\eta_\mathrm{in})\right)
    \nonumber\\
    & + \int_{\eta_\mathrm{in}}^{\eta_0} \md\eta
    \underbrace{\left( 
    \frac{\partial \tilde{\Phi}(\eta,\beta,\hat{k})}{\partial \eta} + 
    \frac{\partial \tilde{\Psi}(\eta,\beta,\hat{k})}{\partial \eta}
    \right)}_{\mathrm{ISW}}
    \Phi_\beta^{~l}\left(\sqrt{|K|}(\eta_0-\eta)\right) 
\end{align}
with $\chi=\sqrt{|K|}(\eta_0-\eta)$. The two contributions on the right-hand side correspond to the so-called ``Sachs-Wolfe'' (SW) effect and ``Integrated-Sachs-Wolfe'' (ISW) effect, respectively. The SW term carries the information of initial anisotropy of GWs $\tilde{\Gamma}_{0}|_{\eta_\mathrm{in}}$ and the gravitational redshift effect from escaping out of the initial potential well $\tilde{\Psi}|_{\eta_\mathrm{in}}$. The ISW term represents the total gravitational redshift effect from traveling along the line of sight.

\section{Angular power spectrum of CGWB and its cross-correlation with CMB}\label{sec:correlations}

In this Section, we will calculate numerically the angular auto-correlation also known as the angular power spectrum of CGWB, which is sensitive to the initial conditions. In the meantime, we will show the cross-correlation of CGWB with CMB temperature anisotropies. At last, we will briefly analyze the signal-to-noise ratio (SNR) for anisotropies from FOPTs.

\subsection{Angular power spectrum}

First of all, we derive in detail the angular power spectrum of CGWB in a curved background. In analogy to the CMB angular power spectrum, the GW anisotropic sky-map $\Gamma(\eta,\vec{x},p,\hat{n})$ can be expanded with spherical harmonics as
\begin{align}\label{eq:Tl_GW}
    \Gamma(\eta,\vec{x},p,\hat{n}) &= \sum_{l,m} a_{lm}^{\mathrm{GW}}(\eta,\vec{x},p) Y_{lm}(\hat{n}), \\
    a_{lm}^{\mathrm{GW}}(\eta,\vec{x},p) &= \int\md^2\hat{n}~ Y_{lm}^*(\hat{n}) \Gamma(\eta,\vec{x},p,\hat{n}) .
\end{align}
The angular power spectrum is defined as the auto-correlation of $a_{lm}^{\mathrm{GW}}$ as $C_l^{\mathrm{GW}} = \left\langle |a_{lm}^{\mathrm{GW}}|^2 \right\rangle$, where the bracket is taken as the ensemble average as below,
\begin{align}
    C_l^{\mathrm{GW}} &= \int\md^2 \hat{n}_1\md^2 \hat{n}_2 Y_{lm}^*(\hat{n}_1)Y_{lm}(\hat{n}_2)
    \int \frac{\md m(\beta_1)}{(2\pi)^3}\frac{\md m(\beta_2)}{(2\pi)^3} \nonumber\\
    &\quad\quad \times \int\md^2\hat{k}_1 \md^2\hat{k}_2 \langle
    \tilde{\Gamma}(\eta,p,\beta_1,\mu_1,\hat{k}_1)
    \tilde{\Gamma}^*(\eta,p,\beta_2,\mu_2,\hat{k}_2)
    \rangle M_{\beta_1}(\hat{k}_1,\chi,\hat{x}) M^*_{\beta_2}(\hat{k}_2,\chi,\hat{x}) \nonumber \\
    &= \sum_{l_1,l_2}(-i)^{l_1 - l_2}(2l_1+1)(2l_2+1) \int \frac{\md m(\beta_1)}{(2\pi)^3}\frac{\md m(\beta_2)}{(2\pi)^3} \nonumber \\
    & \quad\quad \times \int\md^2\hat{k}_1 \md^2\hat{k}_2 \langle\tilde{\Gamma}_{l_1}(\eta,p,\beta_1,\hat{k}_1)
    \tilde{\Gamma}^*_{l_2}(\eta,p,\beta_2,\hat{k}_2)\rangle  M_{\beta_1}(\hat{k}_1,\chi,\hat{x}) M^*_{\beta_2}(\hat{k}_2,\chi,\hat{x}) \nonumber\\
    & \quad\quad \times \int\md^2\hat{n}_1 Y_{lm}^*(\hat{n}_1)P_{l_1}(\mu_1)
    \int\md^2\hat{n}_2 Y_{lm}(\hat{n}_2)P_{l_1}(\mu_2).
\end{align}
It is useful to define a transfer function to relate the $l$-pole fluctuation $\tilde{\Gamma}_{l}$ observed at $\eta$ to the initial perturbations $\tilde{\mathcal{R}}$ at $\eta_\mathrm{in}$,
\begin{align}
    \tilde{\Gamma}_{l}(\eta,p,\beta,\hat{k}) = T_l^\mathrm{GW}(p,\beta,\eta,\eta_\mathrm{in}) \tilde{\mathcal{R}}(\beta,\hat{k}).
\end{align}
Since all the stochastic properties are carried by $\tilde{\Gamma}_{l}$ and $\tilde{\mathcal{R}}$, the transfer function $T_l^\mathrm{GW} =\tilde{\Gamma}_{l}/\tilde{\mathcal{R}}$ should be a deterministic variable rather than a stochastic one, which means that it is just a factor determined by the perturbation equations in FLRW metric and does not participate in the ensemble average. Thus, only $\tilde{\mathcal{R}}$ stays in the ensemble average bracket.

The initial condition for scalar perturbation power spectrum is given by
\begin{align}
    \langle \tilde{\mathcal{R}}(\beta,\hat{k}) \tilde{\mathcal{R}}^*(\beta',\hat{k}') \rangle = (2\pi)^3 \frac{1}{\beta^2}
    \delta(\beta,\beta') \delta^{(2)}(\hat{k} - \hat{k}') P_{\mathcal{R}}(k),
\end{align}
where $k = \sqrt{|K|}\sqrt{\beta^2 - \hat{K}}$ with $\hat{K}\equiv K/|K|$ for $K\neq 0$, and $k=\beta$ with $\hat{K}\equiv 0 $ for $K=0$. The two-point correlation $\langle \mathcal{R}(\vec{x}_1)\mathcal{R}^*(\vec{x}_2)\rangle$ for the initial perturbation as a random field on $M^{(3)}_K$ should be a scalar function, which is independent of the coordinate choice, thus, it can be evaluated by setting $\vec{x}_1$ at origin and $\vec{x}_2$ along $(\theta=0,\phi=0)$ separated by $\chi$,
\begin{align}
    \langle \mathcal{R}(\vec{x}_1)\mathcal{R}^*(\vec{x}_2)\rangle &= \int \frac{\md m(\beta_1)}{(2\pi)^3}\frac{\md m(\beta_2)}{(2\pi)^3}
    \int \md^2\hat{k}_1 \md^2\hat{k}_2  \nonumber\\
    & \quad\quad\quad\quad
    \times \langle \tilde{\mathcal{R}}(\beta_1,\hat{k}_1) \tilde{\mathcal{R}}^*(\beta_2,\hat{k}_2) \rangle
     M_{\beta_1}(\hat{k}_1,0,\hat{x}_1) M^*_{\beta_2}(\hat{k}_2,\chi,\hat{x}_2)
    \nonumber \\
    &= \int \frac{\md m(\beta)}{2\pi^2} P_{\mathcal{R}}(k)
    M_{\beta_1}(\hat{k},0,\hat{x}_1) M^*_{\beta_2}(\hat{k},\chi,\hat{x}_2).
\end{align}
The auto-correlation can be derived by taking the limit $\chi\to 0$,
\begin{align} \label{eq:initial-auto-correlation}
    \langle |\mathcal{R}(\vec{x})|^2\rangle = \int \frac{\md m(\beta)}{2\pi^2} P_{\mathcal{R}}(k) = 
    \int \frac{\md m(\beta)}{\beta^3} \frac{\beta^3}{2\pi^2} P_{\mathcal{R}}(k) \equiv
    \int \frac{\md m(\beta)}{\beta^3} \tilde{\mathcal{P}}_{\mathcal{R}}(k).
\end{align}
Note here that, the dimensionless power spectrum people usually use refers to the power per logarithmic interval of $k$ (denoted as $\mathcal{P}_{\mathcal{R}}(k)$), not per logarithmic interval of $\beta$ (denoted as $\tilde{\mathcal{P}}_{\mathcal{R}}(k)$ in Eq. \eqref{eq:initial-auto-correlation}), which are related to each other by~\cite{Lesgourgues:2013bra}
\begin{align}
    \frac{\md \beta}{\beta} \tilde{\mathcal{P}}_{\mathcal{R}}(k) = 
    \frac{\md k}{k} \mathcal{P}_{\mathcal{R}}(k) \quad \Longrightarrow \quad
    \tilde{\mathcal{P}}_{\mathcal{R}}(k) = \frac{\beta^2}{\beta^2 - \hat{K}}\mathcal{P}_{\mathcal{R}}(k) ~.
\end{align}

Therefore, the angular power spectrum observed today at $\eta=\eta_0$ and $\vec{x}=\vec{x}_0$ can be calculated as
\begin{align}\label{eq:Cl_GW}
    C_l^{\mathrm{GW}} &= (4\pi)^2 \int \frac{\md m(\beta)}{(2\pi)^3} \left|T_l^\mathrm{GW}(p,\beta,\eta_0,\eta_\mathrm{in}) \right|^2 
    P_{\mathcal{R}}(k)
    \int \md^2\hat{k} |Y_{lm}(\hat{k})|^2
    \nonumber \\
    & = 4\pi \int \frac{\md m(\beta)}{\beta(\beta^2 - \hat{K})} \left|T_l^\mathrm{GW}(p,\beta,\eta_0,\eta_\mathrm{in}) \right|^2 \mathcal{P}_{\mathcal{R}}(k),
\end{align}
where we have used the following relations,
\begin{align}
    \int\md^2\hat{n} Y_{lm}^*(\hat{n})P_{l'}(\mu) = \delta_{l,l'} \frac{4\pi}{2l+1}Y_{lm}^*(\hat{k}), \\
    |M_\beta(\hat{k},\chi_0,\hat{x})|^2=1,\quad 
    \int \md^2\hat{k} |Y_{lm}(\hat{k})|^2 = 1.
\end{align}
For a flat space with $K\to 0$, the angular power spectrum \eqref{eq:Cl_GW} returns to the familiar one,
\begin{align}
    C_l^{\mathrm{GW}} &= 4\pi \int_0^\infty \frac{\md k}{k}  \left|T_l^\mathrm{GW}(k,p,\eta_0,\eta_\mathrm{in}) \right|^2 \mathcal{P}_{\mathcal{R}}(k)~.
\end{align}
For later convenience, we further define the more commonly used angular power spectrum, 
\begin{align}\label{eq:Dl_GW}
     D_{l}^{\mathrm{GW}} = \frac{l(l+1) C_l^{\mathrm{GW}}}{2\pi}.
\end{align}

\subsection{Numerical results with different initial conditions}

The transfer function~\eqref{eq:Cl_GW} is directly related to the initial scalar perturbation and the $l$-pole fluctuation~\eqref{eq:Gammal_SW_and_ISW} via definition~\eqref{eq:Tl_GW}. Assuming all the modes of interest are at super-horizon scales, the adiabatic initial condition for relativistic components reads $\tilde{\delta}_{R,\mathrm{in}} = -2\tilde{\Psi}_\mathrm{in}$ with the subscript $R$ for ``Radiation'' and $\tilde{\Psi}_\mathrm{in} = \frac{2}{3}\tilde{R}_\mathrm{in}$. We can then relate $\tilde{\delta}$ with $\tilde{\Gamma}$ as follows. The GW energy spectrum can be evaluated from the distribution function as
\begin{align}
    \Omega_{\mathrm{GW}} = \frac{1}{\rho_c}\frac{\partial \rho_\mathrm{GW}}{\partial \ln p} = \frac{1}{\rho_c}\frac{\partial}{\partial \ln p} \int \md^3\vec{p}~p f_\mathrm{GW} = \frac{p^4}{\rho_c} \int\md^2\hat{n}~f_\mathrm{GW},
\end{align}
where $\rho_c$ denotes the critical density of the Universe today, and $f_\mathrm{GW}$ is given in \eqref{eq:def_Gamma}. The density contrast for GW is given by~\cite{Bartolo:2019oiq,Bartolo:2019yeu}
\begin{align}
    \delta_\mathrm{GW} = \frac{f_{\mathrm{GW}} - \bar{f}_{\mathrm{GW}}}{\bar{f}_{\mathrm{GW}}} = -\frac{\partial\ln\bar{f}_\mathrm{GW}}{\partial\ln p} \Gamma = \left(
        4 -\frac{\partial\ln\bar{\Omega}_\mathrm{GW}}{\partial\ln p}
    \right) \Gamma.
\end{align}

Using the fact that the primordial GW from inflation serves as a radiation-like component of the cosmic fluid at the initial time, it is reasonable to adopt the initial condition $\tilde{\delta}_\mathrm{GW,in} = -2\tilde{\Psi}_\mathrm{in}$. Therefore, the transfer function for PGWB can be normalized by the initial curvature perturbations from Eq.~\eqref{eq:Gammal_SW_and_ISW} as
\begin{align}\label{eq:TSW_l_PGW}
    T_l^\mathrm{SW,PGW} &= \left(1 - \frac{2}{4 - n_{\mathrm{GW}}}\right)\frac{\tilde{\Psi}_\mathrm{in}}{\tilde{\mathcal{R}}_\mathrm{in}} \Phi_\beta^{~l}(\chi_\mathrm{in}), \quad n_{\mathrm{GW}} \equiv \frac{\partial\ln\bar{\Omega}_\mathrm{GW}}{\partial\ln p}, \\
    T_l^\mathrm{ISW,PGW} &= \int_{\eta_\mathrm{in}}^{\eta_0} \md\eta~ \frac{1}{\tilde{\mathcal{R}}_\mathrm{in}}\left( 
    \frac{\partial \tilde{\Phi}}{\partial \eta} + 
    \frac{\partial \tilde{\Psi}}{\partial \eta}
    \right)\Phi_\beta^{~l}(\chi), \quad \chi \equiv \sqrt{|K|}(\eta_0-\eta).
\end{align}
However, the transfer function also gains extra contributions if there is non-adiabatic process induced by the initial non-Gaussianity, which may be significant for scalar induced GWs (SIGWs)~\footnote{There is an alternative treatment for the SIGW case in Ref.~\cite{Dimastrogiovanni:2022eir}.}. In Fourier space, such initial condition reads~\cite{Bartolo:2019zvb,Schulze:2023ich}
\begin{align}
    \tilde{\Gamma}_{0,\mathrm{in}}^{\mathrm{NAD},\mathrm{PGW}} = \frac{3}{5} ~\frac{8f_\mathrm{NL}}{4 - n_{\mathrm{GW}}} \tilde{\mathcal{R}}_\mathrm{in} \equiv \frac{3}{5}\tilde{f}_\mathrm{NL} \tilde{\mathcal{R}}_\mathrm{in},
\end{align}
thus, the corresponding transfer function is given by 
\begin{align}\label{eq:T_NAD}
    T_l^\mathrm{NAD,CGW} &= \frac{3}{5} \tilde{f}_\mathrm{NL}\Phi_\beta^{~l}(\chi_\mathrm{in}).
\end{align}

\begin{figure}[ht]
    \centering
    \includegraphics[width = \textwidth]{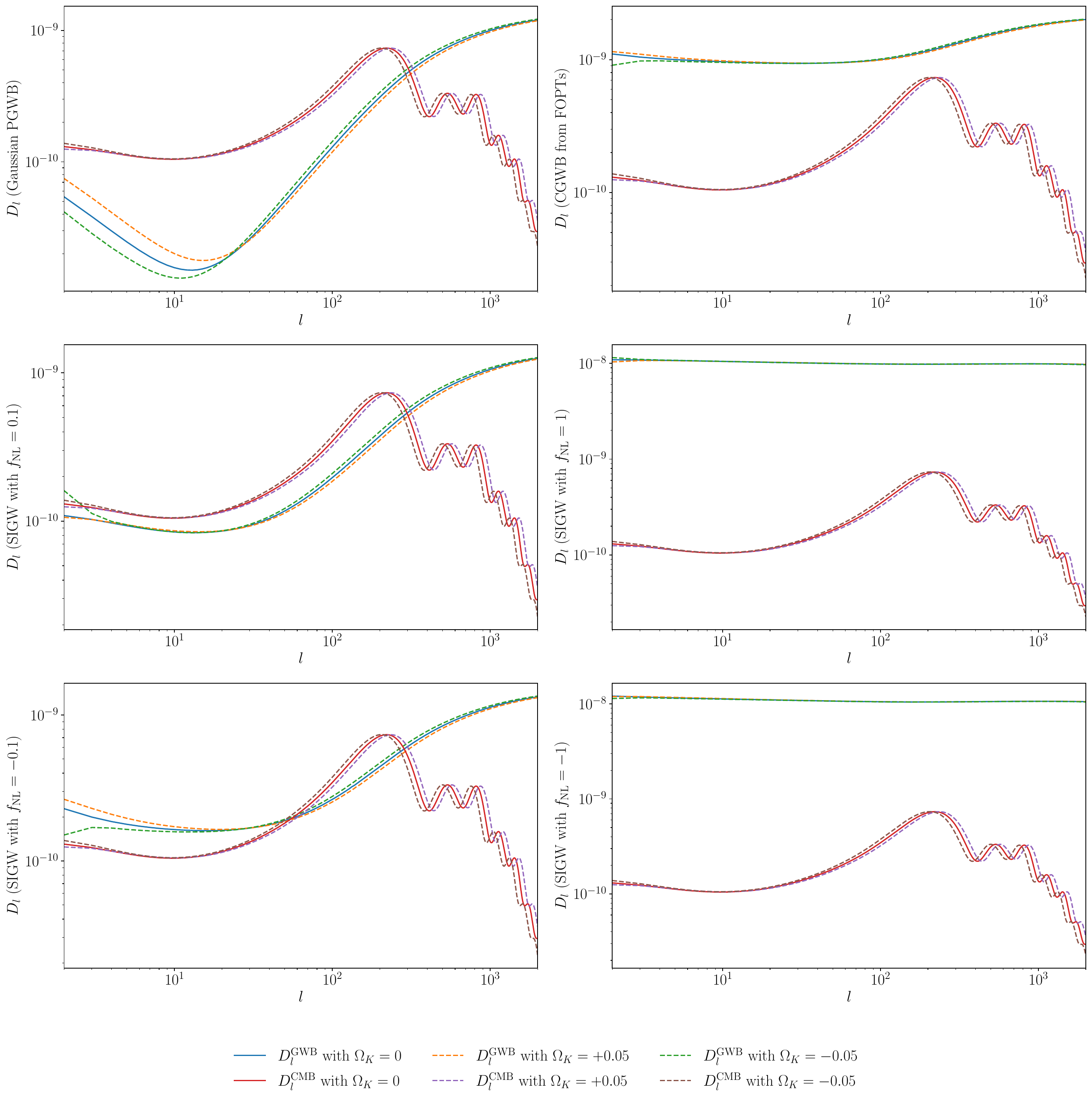}
    \caption{The comparison of angular power spectra $D_l\equiv l(l+1)C_l/(2\pi)$ between CMB and GWB for Gaussian PGWB, GWB from FOPTs, GWB with $f_\mathrm{NL}=0.1$, $f_\mathrm{NL}=1$, $f_\mathrm{NL}=-0.1$, and $f_\mathrm{NL}=-1$, respectively.}
    \label{fig:angular_power_spectrum}
\end{figure}

However, stories get changed for GWs generated from the cosmic fluid motions, for example, the cosmological FOPTs, where the GW energy density posses a squared dependence on the stress tensor of the cosmic fluid~\cite{Hindmarsh:2016lnk,Jinno:2016knw,Cai:2023guc,Guo:2020grp},
\begin{align}
    \rho_\mathrm{GW} \propto \Lambda_{ij,kl}\Lambda_{ij,mn} T_{kl} T_{mn} \propto \rho_{f}^2.
\end{align}
To the leading order, the perturbations of energy density are related by
\begin{align}
    \rho_\mathrm{GW}(1+\delta_\mathrm{GW}) \propto \rho_{f}^2(1+\delta_f)^2 \simeq \rho_{f}^2(1+2\delta_f), \quad \Rightarrow \quad \delta_\mathrm{GW} = 2 \delta_f.
\end{align}
For most of the FOPTs we are interested in, such as the electro-weak PT and QCD PT, the cosmic fluid is mainly made of radiation-like fluids, and all the modes are well outside the horizon. Therefore, we have $\tilde{\delta}_{GW} = 2\tilde{\delta}_{f} = 2\tilde{\delta}_{R,\mathrm{in}} = -4\tilde{\Psi}_\mathrm{in}$, which leads to the following transfer functions,
\begin{align}\label{eq:TSW_l_PT}
    T_l^\mathrm{SW,PT} &= \left(1 - \frac{4}{4 - n_{\mathrm{GW}}}\right)\frac{\tilde{\Psi}_\mathrm{in}}{\tilde{\mathcal{R}}_\mathrm{in}} \Phi_\beta^{~l}(\chi_\mathrm{in}), \\
    T_l^\mathrm{ISW,PT} &= \int_{\eta_\mathrm{in}}^{\eta_0} \md\eta~\frac{1}{\tilde{\mathcal{R}}_\mathrm{in}}\left( 
    \frac{\partial \tilde{\Phi}}{\partial \eta} + 
    \frac{\partial \tilde{\Psi}}{\partial \eta}
    \right)\Phi_\beta^{~l}(\chi).
\end{align}
Here we do not consider non-Gaussianity contributions in GWs from FOPTs. As we can see, the ISW transfer function remains unchanged since it describes the gravitational redshift along the traveling 
path, but the SW transfer function gains a different factor to reflect different mechanisms.

Several Boltzmann codes have already been made public, including CLASS~\cite{Lesgourgues:2011re} and CAMB~\cite{Lewis:1999bs}, and one another code worth emphasizing is GW\textunderscore CLASS~\cite{Schulze:2023ich}, which is a public code dealing with CGWB anisotropies . In this work, we have modified the CAMB code to numerically calculate the angular power spectrum $C_{l}^{\mathrm{GWB}}$ and the cross-correlations between CGWB and CMB for a general $\Omega_K$. The amplitude and spectral index of power-law power spectrum from primordial curvature perturbations are fixed by $\ln(10^{10}A_{s}) = 3.04$ and $n_s=0.966$ at the standard pivot scale $k_{\mathrm{pivot}} = 0.05 \mathrm{Mpc}^{-1}$. For GWs at nano-Hertz band, the PTA data indicates that one can parameterize the GW background with a power-law spectrum $\Omega_{\mathrm{GW}}h^2 \sim (f/\mathrm{Hz})^{n_{\mathrm{GW}}}$ with $n_{\mathrm{GW}} \simeq 1.82$~\cite{Ding:2023xeg}. In Ref.~\cite{DiValentino:2019qzk}, the authors found that a closed Universe with $\Omega_{K}=-0.0438$ gives a better fit to Planck 2018 data concerning the standard flat model. Therefore, we will illustrate with $\Omega_{K}=\pm 0.05$ in our numerical results. The auto-correlation angular power spectra of CGWB with different initial conditions discussed above, like the Gaussian PGWB, GWB from FOPTs, GWB with $f_\mathrm{NL}=0.1, 1,-0.1,-1$ in the flat (blue solid), open (orange dashed), and closed (green dashed) Universe are shown in Fig.~\ref{fig:angular_power_spectrum}, along with their comparison with the CMB angular power spectrum in the flat (red solid), open (purple dashed), and closed (purple dashed) Universe. Since GWs decoupled from the cosmic fluid almost at the moment of their generations, there is no Silk-damping analog at small scales like CMB, resulting in a plateau at large $l$.

\subsection{Cross-correlations between CGWB and CMB}

Since gravitons and CMB photons share the same perturbations, they have the same geodesics during their propagation, which leads to the cross-correlations between CGWB and CMB~\cite{Malhotra:2020ket,Adshead:2020bji,Ricciardone:2021kel}. After a similar procedure, one can obtain the perturbation in the distribution function of CMB photon as
\begin{align}\label{eq:Theta_l_CMB}
    \Theta_{l}(\beta,\eta_{*},\eta_{0}) =& \int_{\eta_{*}}^{\eta_{0}}\md\eta \bigg\{ g(\eta)\left( \frac{1}{4}\tilde{\delta}_\gamma(\beta,\eta) + \Psi(\beta, \eta)  \right)\Phi_{\beta}^{~l}(\chi) + g(\eta)\frac{-i\tilde{v}_{b}(k, \eta)\sqrt{|K|}}{k^2} \frac{\md\Phi_{\beta}^{~l}(\chi)}{\md\chi}\nonumber\\
    & \quad + e^{-\tau} \left( \tilde{\Psi}^{\prime}(\beta, \eta) + \tilde{\Phi}^{\prime}(\beta, \eta) \right)\Phi_{\beta}^{~l}(\chi) \bigg\} \nonumber\\
    \simeq & \underbrace{\left(\Theta_{0}(\beta,\eta_{*}) + \Psi(\beta, \eta_{*})\right)\Phi_{\beta}^{~l}(\chi_*)}_{\mathrm{SW}} + \underbrace{\int_{\eta_{*}}^{\eta_{0}}\md\eta \left(\tilde{\Psi}^{\prime}(\beta, \eta) + \tilde{\Phi}^{\prime}(\beta, \eta)\right)\Phi_{\beta}^{~l}(\chi)}_{\mathrm{ISW}} \nonumber\\
    &+ \underbrace{3\Theta_1(\beta,\eta_*) \left(
        \left(\frac{l+1}{2l+1} \frac{s_l}{s_{l+1}} + \frac{l}{2l+1}\right) \Phi_\beta^{~l-1}(\chi_*) - \frac{(l+1)\cot_{\mathrm{K}}\chi_*}{\sqrt{\beta^2 - \hat{K}(l+1)^2}} \Phi_\beta^{~l}(\chi_*)
    \right)}_{\mathrm{Doppler}},
\end{align}
where $v_{b}$ is the velocity of baryons, $\tau(\eta) = \int_{\eta}^{\eta_{0}}\md \eta^{\prime}n_{e}\sigma_{T}a$ is the optical depth, and $g(\eta) = -\tau^{\prime}(\eta)e^{-\tau(\eta)}$ is the visibility function. To arrive at the second line, we have used the instantaneous recombination assumptions $g(\eta) = \delta(\eta - \eta_{*})$ and $e^{-\tau} = \theta(\eta - \eta_{*})$.
The cross-correlation between GWB and CMB is defined as
\begin{align}
    C_{l}^{\mathrm{GW}\times\mathrm{CMB}} \equiv 4\pi \int \frac{\md m(\beta)}{\beta(\beta^2 - \hat{K})} T_l^\mathrm{GW}(p,\beta,\eta_0,\eta_\mathrm{in}) \Theta_l(\beta,\eta_0,\eta_*) \mathcal{P}_{\mathcal{R}}(k).
\end{align}

\begin{figure}[t]
    \centering
    \includegraphics[width =\textwidth]{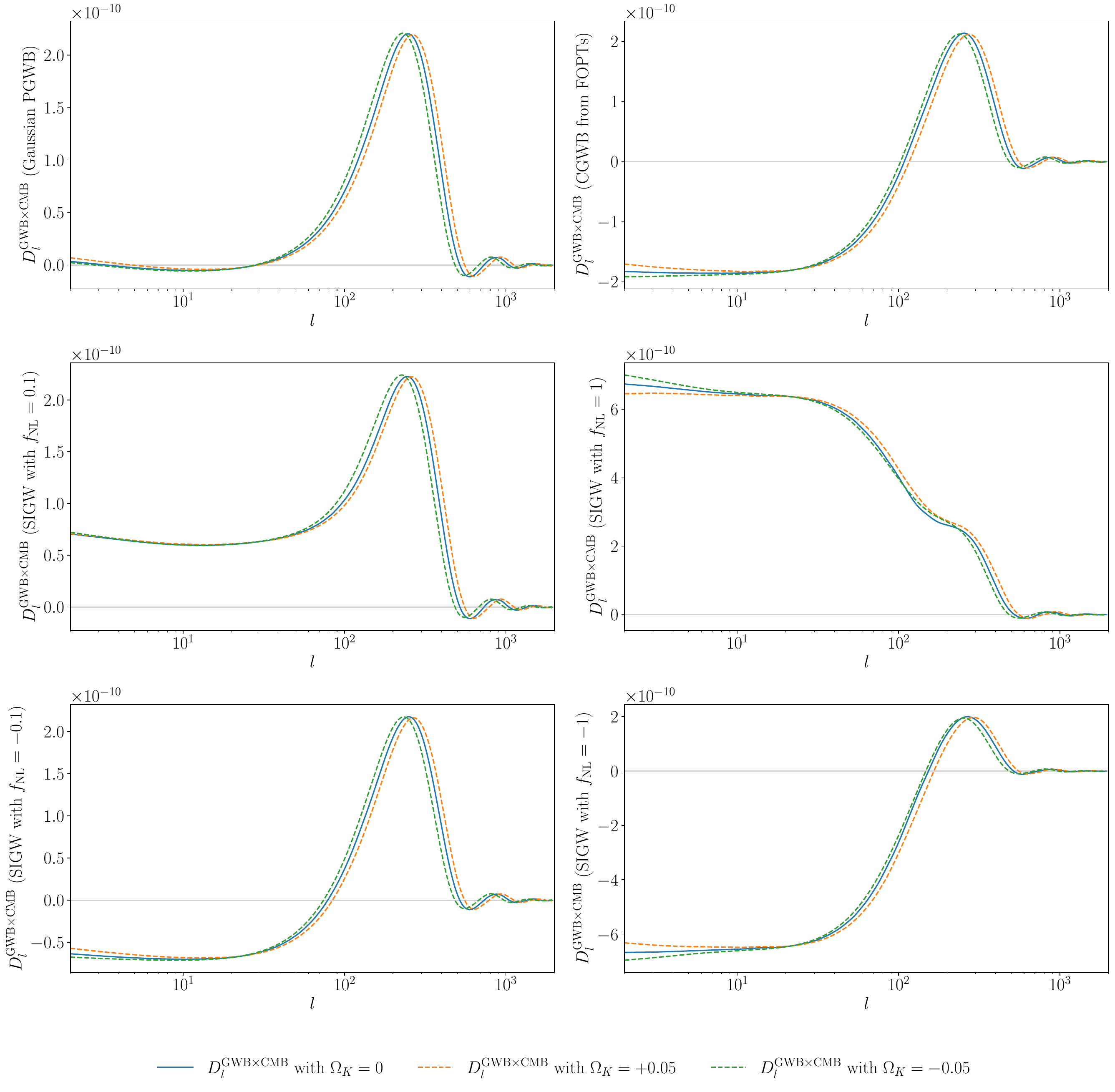}
    \caption{The cross-correlation angular power spectra $D_l\equiv l(l+1)C_l/(2\pi)$ between CMB and GWB in the flat (blue solid), open (orange dashed), and closed (green dashed) Universe, where GWB spectra with different initial conditions are presented from the Gaussian PGWB, GWB from FOPTs, PGWB with $f_\mathrm{NL}=0.1$, $f_\mathrm{NL}=1$, $f_\mathrm{NL}=-0.1$, and $f_\mathrm{NL}=-1$, respectively.}
    \label{fig:cross-correlation}
\end{figure}

The cross-correlations between CMB temperature anisotropy and GWB with different initial conditions are shown in Fig.~\ref{fig:cross-correlation}. The acoustic peaks from CMB leave clear imprints on the cross-correlations and deviate slightly for different spatial geometries. Due to the silk-damping in CMB, the cross-correlation rapidly decays to zero at small scales (large $l$). For the case like the Gaussian PGWB where GWB anisotropy is dominated by the ISW effect, the cross-correlation at large scales (small $l$) almost vanishes since CMB temperature anisotropy is dominated by the SW effect. For the case where GWB anisotropy is dominated by the SW effect, an SW plateau similar to the CMB angular power spectrum arises at the large-scale cross-correlations. However, the plateau may be greater or less than zero, indicating a positive or negative correlation between CMB and GWB. For GWs from FOPTs, the signs of SW effects in GWB and CMB are different, leading to a negative correlation. For PGWB with non-Gaussianity, the SW effect is much smaller than the non-adiabatic contribution, which shares the same form (proportional to hyper-spherical Bessel function) with the SW effect in CMB. Therefore, the main contribution to the plateau of PGWB with $f_\mathrm{NL}\neq 0$ comes from the cross-correlation between non-adiabatic transfer function~\eqref{eq:T_NAD} and the SW effect~\eqref{eq:Theta_l_CMB}, the value of which is positive for $f_{\mathrm{NL}}>0$ and is negative for $f_{\mathrm{NL}}<0$.

It should be kept in mind that the contribution from the CGWB monopole heavily depends on $n_\mathrm{GW}$, where $n_\mathrm{GW}\simeq 2$ strongly suppresses the SW effect \eqref{eq:TSW_l_PGW} for PGWB but keeps a finite value for GWs from FOPTs~\eqref{eq:TSW_l_PT}. Specifically, the preferred value $n_\mathrm{GW}\simeq 1.82$ from PTA data results in a strong cancellation between $\tilde{\Gamma}_{0,\mathrm{in}}$ and $\tilde{\Psi}_\mathrm{in}$ for PGWs but not for GWs from FOPTs, leading to significantly different amplitudes in auto-correlation and cross-correlation.

\subsection{SNR of anisotropies from FOPTs}

\begin{figure}[t]
    \centering
    \includegraphics[width =0.95\textwidth]{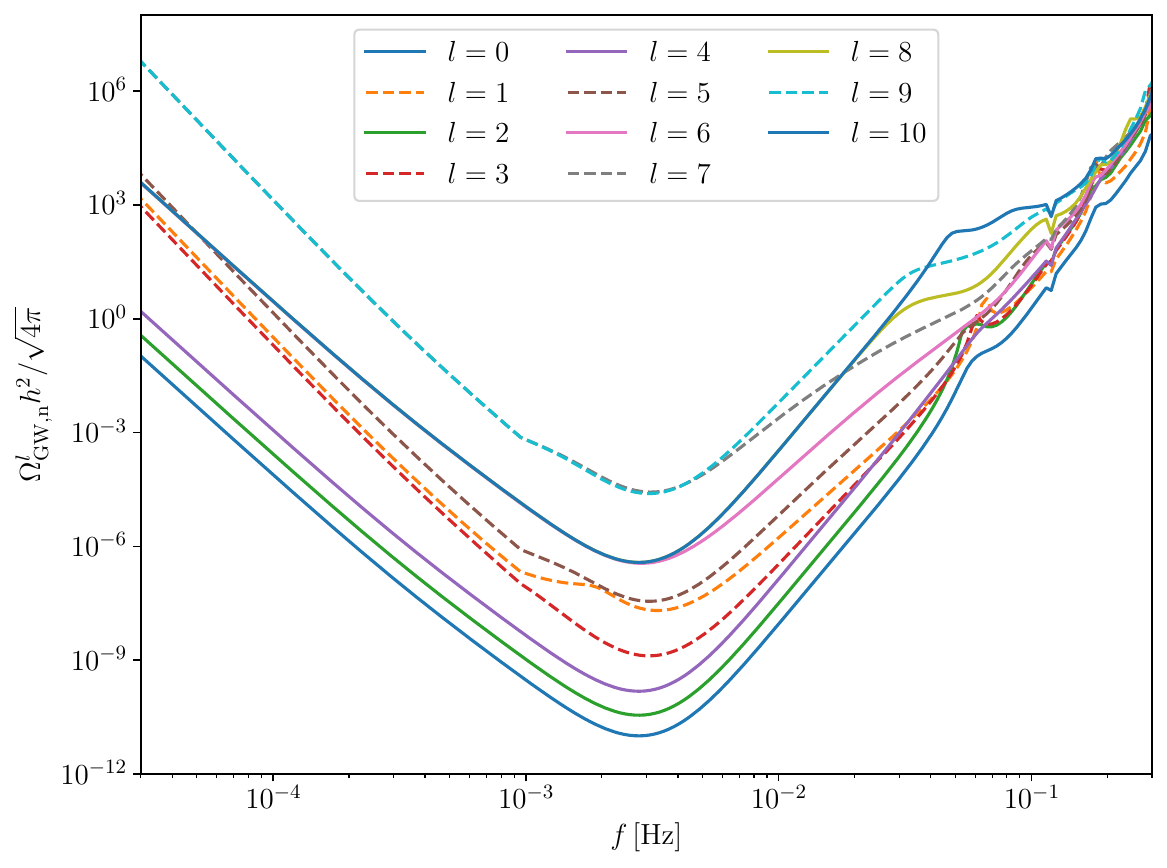}
    \caption{The effective sensitivity of the detector network LISA, for multipoles up to $l=10$. Even (odd) multipoles are shown with solid (dashed) lines.}
    \label{fig:nl_LISA}
\end{figure}

To relate to the experiments, an analysis of SNR is needed. We will follow the definition of Refs.~\cite{LISACosmologyWorkingGroup:2022kbp,Dimastrogiovanni:2022eir,Alonso:2020rar}, and the SNR is related to the multipole $\mathrm{SNR}_{l}^2$ via $\mathrm{SNR}^2 =\sum_{l}(2l+1)\mathrm{SNR}_{l}^2$, where the multipole SNR, $\mathrm{SNR}_{l}$, is defined as
\begin{align}\label{eq:SNRl}
    \mathrm{SNR}_{l}^2 = \int \mathbf{d}f C_{l}^{\mathrm{GW}}(f)
    \left( \frac{\Omega_{\mathrm{GW}}(f)}{\Omega_{\mathrm{GW},n}^{l}(f)}  \right)^2.
\end{align}
Here, $\Omega_{\mathrm{GW},n}^l(f)$ is the effective angular sensitivity of the detector network to the $l$-th multipole,
\begin{align}
    \Omega_{\mathrm{GW},n}^l(f)^{-2} \equiv T_{\mathrm{obs}}\sum_{AB}\left(\frac{2}{5}\right)^2 \left(\frac{4\pi^2f^3}{3H_{0}^2}\right)^{-2}\frac{\sum_{m}|\mathcal{A}_{AB}^{lm}{f}|^2}{N_{f}^AN_{f}^B(2l+1)},
\end{align}
where $T_{\mathrm{obs}}$ is the time of observation, $N_{f}^A$ is the noise power spectrum density for detector A, and $\mathcal{A}_{AB}^{lm}$ denotes the spherical harmonic transform of the antenna pattern for the detector pair $AB$.
In the following, we will use the code $\textbf{schNell}$~\cite{Alonso:2020rar} to calculate $\Omega_{\mathrm{GW},n}^l(f)$. As for the numerical part, we choose the observation time $T_{\mathrm{obs}}$ to be 4 years and fix the pivot frequency to be 1 mHz in the code. We adopt the effective sensitivity of the detector network LISA~\footnote{This result is marginally consistent with the analysis of Ref.~\cite{LISACosmologyWorkingGroup:2022kbp}, where the discontinuity of $l=1$ curve around 3 mHz might be numerical instability but it is unharmful to our calculations after integrations.} as shown in Fig.~\ref{fig:nl_LISA}, and one can see the sensitivity declines quickly when $l$ gets larger.

\begin{figure}[t]
    \centering
    \includegraphics[width =\textwidth]{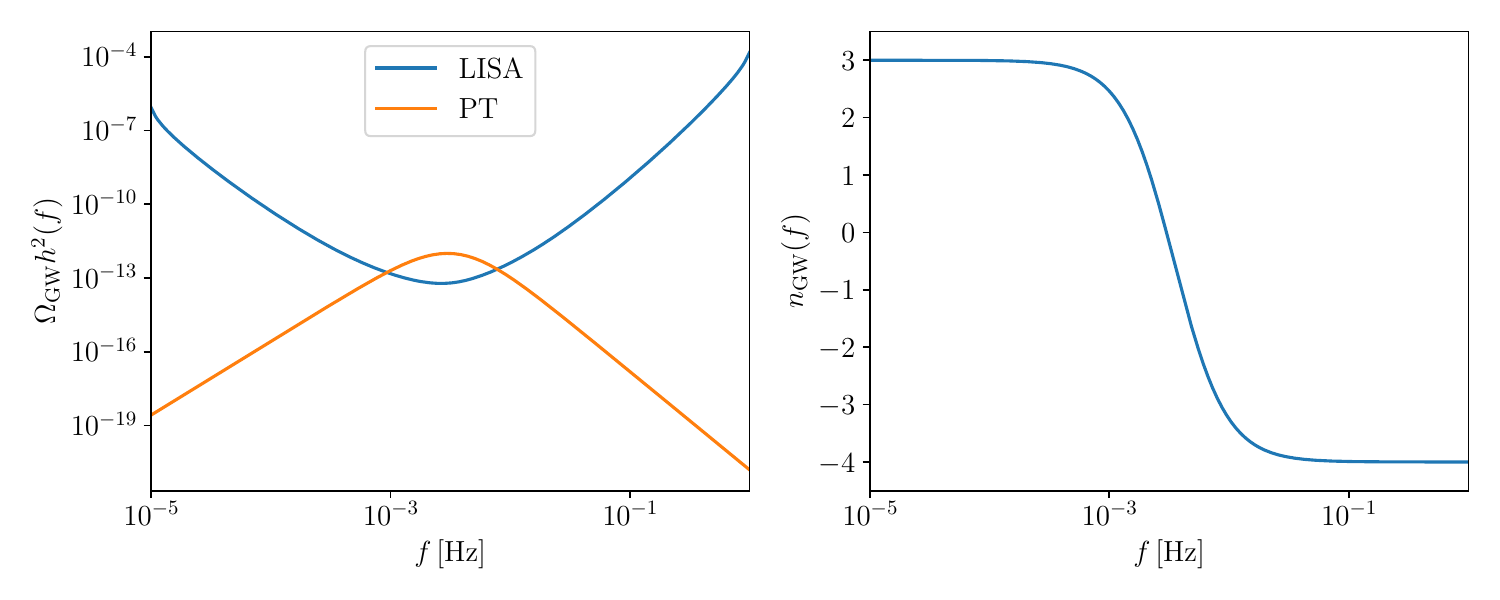}
    \caption{\textit{Left}: The gravitational wave energy density spectrum of phase transition, and the sensitive curve of LISA detector is also given. The amplitude $A$ is chosen as $A=10^{-12}$, which is consistent with the numerical result of~\cite{Hindmarsh:2017gnf,Guo:2020grp}, and the peak frequency is set to be $f_{p} = 3\times 10^{-3}$Hz. \textit{Right}: The tensor spectrum tilde $n_{\mathrm{Gw}}$ is given, and one can see it depends on frequency in this situation. If the peak frequency shift to other frequency band, one will get a constant $n_{\mathrm{Gw}}$ in this $\mathcal{O}$(mHz) frequency band.}
    \label{fig:GW_PT}
\end{figure}

Recalling that although our analysis above this subsection is general, we have specifically adopted an illutrative value for $n_{\mathrm{GW}}$~\eqref{eq:TSW_l_PGW} from the recent fitting result of PTA data~\cite{Ding:2023xeg}, and hence all numerical results above this subsection are subjected to the frequency band of $\mathcal{O}(\mathrm{nHz})$~\footnote{For those who are interested in the PTA sensitivity of anisotropies, please refer to Refs.~\cite{Depta:2024ykq,NANOGrav:2023tcn}, and for those who are interested in other detectors or frequency bands, please refer to Ref.~\cite{Cui:2023dlo}.}. Nevertheless, our analysis can easily be extended to the frequency band of $\mathcal{O}(\mathrm{mHz})$ given a specific model of GW energy spectrum at that frequency band. In this subsection, we will roughly estimate the SNR for the case with a FOPT using the LISA detector network, while generalizing to other cases is straightforward. As for the phase-transition case, we consider the total GW spectrum $\Omega_{\mathrm{GW}}$ is dominated by the sound-wave spectrum $\Omega_{\mathrm{sw}}$ fitted by the LISA Cosmology Working Group (see, for example, Ref.~\cite{Caprini:2015zlo}),
\begin{align}
    \Omega_{\mathrm{GW}}^\mathrm{PT}(f)h^2 = AS_{\mathrm{sw}}(f/f_{p}),
\end{align}
where $f_p$ refers to the peak frequency, and the shape function is fitted as
\begin{align}
    S_{\mathrm{sw}}(f/f_{p}) \equiv \left( \frac{f}{f_p} \right)^3 \left( \frac{7}{4+3(f/f_p)^2} \right)^{7/2}.
\end{align}
We will choose $A=10^{-12}$ to be consistent with numerical results from Refs.~\cite{Hindmarsh:2017gnf,Guo:2020grp}, and the peak frequency is fixed at $f_{p} = 3\times10^{-3}$Hz, which is shown in the left panel of Fig.~\ref{fig:GW_PT}. In this case, $n_{\mathrm{GW}}$ is a frequency-dependent quantity as shown in the right panel of Fig.~\ref{fig:GW_PT}, which asymptotes to constant powers at lower and higher frequencies.

\begin{figure}[t]
    \centering
    \includegraphics[width =0.9\textwidth]{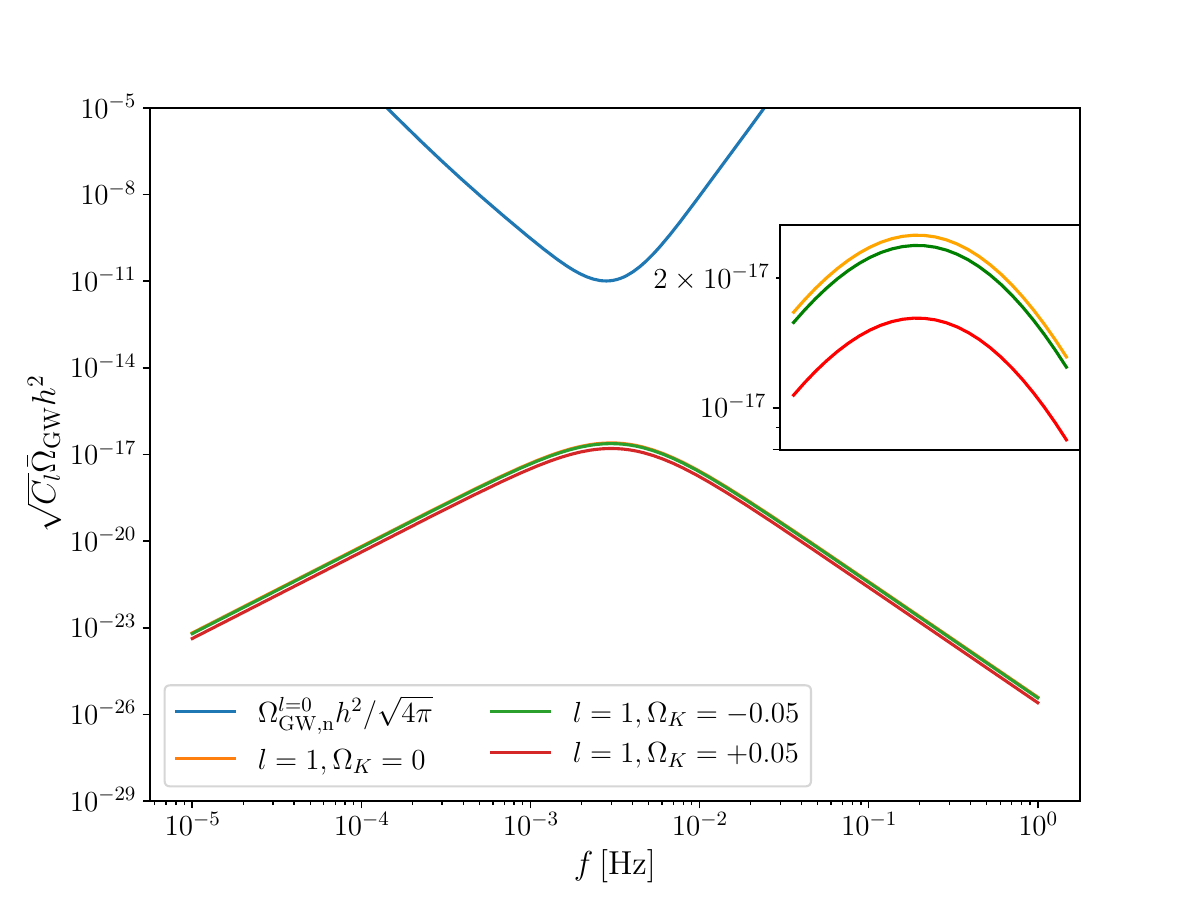}
    \caption{The strength of PT signal $\sqrt{C}_{l}\bar{\Omega}_{\mathrm{GW}}h^2$ and the sensitivity curve of LISA detector.}
    \label{fig:S_PT}
\end{figure}

\begin{figure}[t]
    \centering
    \includegraphics[width =\textwidth]{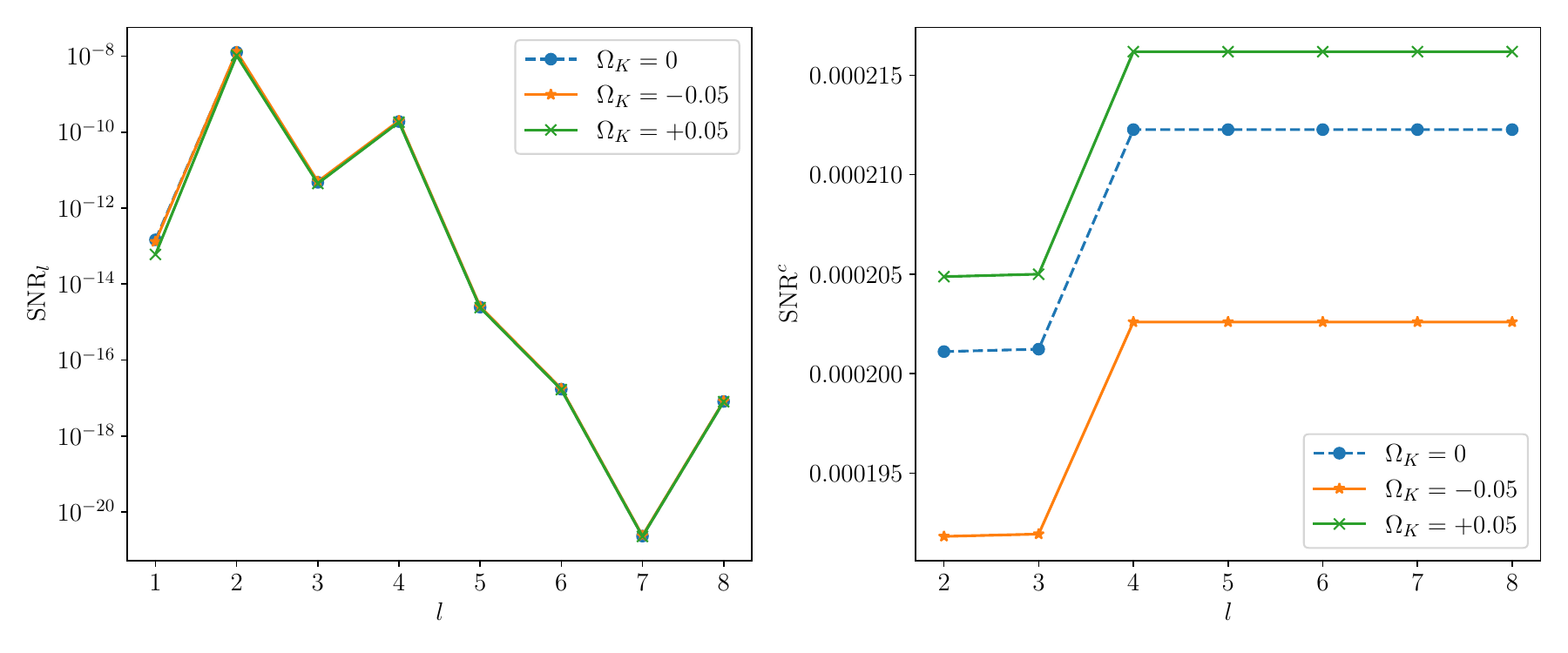}
    \caption{\textit{Left}: The $\mathrm{SNR}_{l}$ of the phase transition mechanism. \textit{Right}: The $\mathrm{SNR}^c$ of the cross-correlation in the phase transition case.}
    \label{fig:SNR}
\end{figure}

Before we calculate $\mathrm{SNR}_l$, we can visualize the relative magnitudes of $\Omega_\mathrm{GW}^\mathrm{PT}$ with respect to the most optimistic sensitivity $\Omega_\mathrm{GW}^{l=0}$ as shown in Fig.~\ref{fig:S_PT}, where we have multiplied the monopole part of $\Omega_\mathrm{GW}$ with $\sqrt{C_l}$ as appear in the integrand of definition of SNR~\eqref{eq:SNRl}. The numerical result of each $\mathrm{SNR}_l$ is presented in the left panel of Fig.~\ref{fig:SNR} for different choices of spatial curvature, which is far below the current detection ability of LISA. To obtain $\mathcal{O}(1)$ SNR, we can raise the amplitude $A$ of $\bar{\Omega}_{\mathrm{GW}}$ up to $\mathcal{O}(10^{-6})$. Therefore, it will be quite challenging for LISA to detect the corresponding SNR, which motivates us to look at the cross-correlation, whose SNR is defined as~\cite{Ricciardone:2021kel}
\begin{align}
    (\mathrm{SNR}^{c})^2 = f_{\mathrm{sky}}\sum_{l=2}^{l_{\mathrm{max}}} (2l+1)
    \frac{(C_{l}^{\mathrm{CMB}\times\mathrm{GWB}})^2}{(C_{l}^{\mathrm{CMB}\times\mathrm{GWB}})^2 + (C_{l}^{\mathrm{GWB}}+N_l^{\mathrm{GWB}})C_{l}^{\mathrm{CMB}}}
    \equiv f_{\mathrm{sky}}\sum_{l=2}^{l_{\mathrm{max}}} (2l+1)(\mathrm{SNR}_{l}^{c})^2.
\end{align}
Here, $f_{\mathrm{sky}}$ denotes the sky coverage that can be simply set to $1$ for a full-sky coverage, and we also neglect the noise of CMB instruments since it is much smaller than $C_{l}^{\mathrm{CMB}}$ at low $l$. The $l$-th multipole noise power spectrum $N_l^{\mathrm{GWB}}$ can be computed as~\footnote{Note the cosmic variance is not included for both $N_l^{\mathrm{GWB}}$ and $N_{l}^{\mathrm{CMB}}$ for an order-of-magnitude estimation.}
\begin{align}
    N_{l}^{\mathrm{GWB}} = \left[\int \mathrm{d}f \frac{\bar{\Omega}_{\mathrm{GW}}(f)^2}{\Omega_{\mathrm{GW,n}}^l(f)^2}  \right]^{-1}.
\end{align}
In the right panel of Fig.~\ref{fig:SNR}, we have shown the numerical results of $\mathrm{SNR}^{c}$, which is much larger than the $\mathrm{SNR}$ of auto-correlation. Again, if the amplitude $A$ of the GW power spectrum can reach up to $\mathcal{O}(10^{-9})$, the SNR of cross-correlation would arrive at $\mathcal{O}(1)$, which is consistent with Ref.~\cite{Ricciardone:2021kel}.

\section{Conclusion and discussion}\label{sec:condis}

Due to the extremely weak interactions of gravitons with ordinary matters, the CGWB serves as a clean probe into the early Universe. Furthermore, the anisotropies in the CGWB can offer additional valuable insights into various physical mechanisms generating the CGWBs, such as alleviating the degeneracy of $F_{\mathrm{NL}}$ in scalar-induced gravitational waves (SIGW)~\cite{Li:2023qua}. In this study, we have extended the analysis of CGWB anisotropies to a non-flat background. In addition to the conventional total angular momentum method, a more direct approach is proposed to perform harmonic decompositions on scalar functions and then address the Boltzmann equations in a non-zero spatial curvature background with minimal assumptions. We applied this method to derive the expression of auto-correlation and cross-correlation of GWB and CMB angular power spectra. We have shown that variations in initial conditions can significantly influence the final angular power spectra of different CGWBs, as these conditions can either enhance or suppress the SW effect while keeping the ISW effect the same as each other, which can lead to different behaviors when spatial curvature is non-zero. Therefore it is important to study the case when the initial condition is non-adiabatic~\cite{ValbusaDallArmi:2023nqn} or isocurvature~\cite{Malhotra:2022ply}, which may contribute a very different SW effect. We have also prospected the detection ability of LISA for the GW anisotropies from a FOPT.

Additionally, relativistic particles may also impact the results~\cite{ValbusaDallArmi:2020ifo} due to their effects on scalar perturbations when the anisotropic perturbation is non-zero (namely $\Phi \neq \Psi$), which can subtly alter the SW and ISW effects, otherwise, the result will not change~\cite{Malhotra:2022ply}. Hence, a different number of relativistic particles can also affect the final angular power spectrum when the spatial curvature is non-zero. For the SIGW case, we only consider in this paper the situation calculated in Ref.~\cite{Bartolo:2019zvb}, therefore, some other cases like Ref.~\cite{Dimastrogiovanni:2022eir} need to be considered separately. Furthermore, a recent study~\cite{Kumar:2024bvp} on SIGW in non-flat spacetime has indicated a very different shape in the GW energy density spectrum solely due to the curvature effect. As a result, a more systematic analysis of the SIGW case is needed, which will be explored in future studies.

\acknowledgments

This work is supported by 
the National Key Research and Development Program of China Grants No. 2021YFC2203004, No. 2021YFA0718304, and No. 2020YFC2201501,
the National Natural Science Foundation of China Grants No. 12422502, No. 12105344, No. 12235019,  No. 12047503, No. 12073088, No. 11821505, No. 11991052, and No. 11947302,
the Strategic Priority Research Program of the Chinese Academy of Sciences (CAS) Grant No. XDB23030100, No. XDA15020701, the Key Research Program of the CAS Grant No. XDPB15,  the Key Research Program of Frontier Sciences of CAS,
the Science Research Grants from the China Manned Space Project with No. CMS-CSST-2021-B01 (supported by China Manned Space Program through its Space Application System). Our code is available at this \href{https://github.com/Einste11N/GW_Aniso_With_K/tree/main}{ Github repository}.

\appendix

\section{Hierarchy equations}\label{app:hierarchy}

In this Appendix, we show a detailed derivation of the hierarchy equations. Directly performing multipole decomposition on Eq.~\eqref{eq:Boltzmann_for_Gamma_tilde} results in
\begin{align}\label{eq:Boltzmann_decomposed}
    \frac{i^l}{2}\int_{-1}^1\md\mu P_l(\mu) \left(
        \frac{\partial \tilde{\Gamma}}{\partial \eta}M_\beta + a \hat{n}^i\tilde{\Gamma} \partial_i M_\beta -  \frac{\partial \tilde{\Phi}}{\partial \eta} M_\beta + a \hat{n}^i\tilde{\Psi} \partial_i M_\beta
    \right) = 0~,
\end{align}
where each term in the integrand is presented as a product of two functions of $\mu$. For any individual function of $\mu$, the multipole decomposition can be applied. The product of two functions $A(\mu)$ and $B(\mu)$ can be decomposed as a single function of $\mu$,
\begin{align}
     & A(\mu)B(\mu) \equiv F(\mu) =\sum_{l}(-i)^{l}(2l_1+1) P_{l}(\mu) F_{l}~, \quad F_{l} = \frac{i^l}{2}\int_{-1}^{1}\md\mu~P_l(\mu) A(\mu)B(\mu)~.
\end{align}
It can also be expressed by performing multipole decomposition separately on $A$ and $B$,
\begin{align}
    A(\mu)B(\mu) = \sum_{l_1}(-i)^{l_1}(2l_1+1) P_{l_1}(\mu) A_{l_1} 
    \sum_{l_2}(-i)^{l_2}(2l_2+1) P_{l_2}(\mu) B_{l_2}~,
\end{align}
which should be equivalent to the first one. Indeed, with the Clebsch-Gordan relation~\cite{Hu:1997hp}
\begin{align}
    P_{l_1}(\mu) P_{l_2}(\mu) = \sum_{l} (2l+1) 
    \left(\begin{array}{ccc}
        l_1 & l_2 & l \\
        0 & 0 & 0
    \end{array}\right)^2 P_l(\mu),
\end{align}
one can relate $F_l$ with $A_l$ and $B_l$ by
\begin{align}
    F_l = \sum_{l_1,l_2} (-i)^{l_1+l_2-l} (2l_1+1)(2l_2+1)  \left(\begin{array}{ccc}
        l_1 & l_2 & l \\
        0 & 0 & 0
    \end{array}\right)^2 A_{l_1} B_{l_2}~.
\end{align}
Then, one can use this relation to rewrite each term in Eq.~\eqref{eq:Boltzmann_decomposed} as a summation over $l_1$ and $l_2$. For example, the first term in the integrand can be recast into
\begin{align}
    \frac{i^l}{2}\int_{-1}^{1}\md\mu~P_l(\mu) \frac{\partial \tilde{\Gamma}}{\partial \eta}M_\beta = \sum_{l_1, l_2}(-i)^{l_1 + l_2} (2l_1 + 1)(2{l_2}+1) 
    \left(\begin{array}{ccc}
        l_1 & l_2 & l \\
        0 & 0 & 0
    \end{array}\right)^2
    \frac{\partial \tilde{\Gamma}_{l_1}}{\partial \eta} \Phi_\beta^{~{l_2}}~,
\end{align}
where we have used the fact
\begin{align}
    \tilde{\Gamma} = \sum_{l_1} (-i)^{l_1} (2{l_1}+1)P_{l_1}(\mu) \tilde{\Gamma}_{l_1} ~,\quad
    M_\beta = \sum_{l_2} (-i)^{l_2} (2l_2+1)P_{l_2}(\mu) \Phi_\beta^{~l_2}~.
\end{align}
The same procedure can be applied to the remaining three terms, which requires the multipole expansions of each factor. For example, the expansions of $\tilde{\Phi}$ and $\tilde{\Psi}$ are trivial since they do not depend on $\mu$ and thus there are only monopole contributions,
\begin{align}
     \tilde{\Phi}_l = \tilde{\Phi} \delta_{l,0}~,\quad \tilde{\Psi}_l = \tilde{\Psi} \delta_{l,0}~.
\end{align}
As for the expansion of $a \hat{n}^i \partial_i M_\beta$, using Eq.~\eqref{eq:M_beta_of_mu}, the radial derivative of $M_\beta$ is given by
\begin{align}
    a \hat{n}^i \partial_i M_\beta &= -\sqrt{|K|}\frac{\partial  M_\beta}{\partial\chi} = -\sum_{l} (-i)^l (2l+1) \frac{\partial \Phi_\beta^{~l}}{\partial\chi}P_l(\mu) \nonumber\\
    &= \sum_{l} (-i)^l (2l+1) \left( \frac{(l+1) s_{l+1}}{2l+1}\Phi_\beta^{~l+1} - \frac{l s_l}{2l+1}\Phi_\beta^{~l-1}\right) P_l(\mu),
\end{align}
where we have used the recurrence relations~\cite{Abbott:1986ct},
\begin{align}
    \Phi_\beta^{~l} &= \frac{1}{\sqrt{\beta^2 - \hat{K} l^2}} \left((2l-1)\cot_\mathrm{K}\chi~\Phi_\beta^{~l-1}  - \sqrt{\beta^2 - \hat{K} (l-1)^2}~\Phi_\beta^{~l-2}\right), \\
     \frac{\partial \Phi_\beta^{~l}}{\partial\chi} &= l \cot_\mathrm{K}\chi~\Phi_\beta^{~l} - \sqrt{\beta^2 - \hat{K} (l+1)^2}~\Phi_\beta^{~l+1},
\end{align}
with an abbreviation $s_{l} \equiv \sqrt{|K|} \sqrt{\beta^2 - \hat{K}l^2}$ for $K\neq 0$ and $s_l\equiv \beta$ for $K=0$.

Plugging all the results above in Eq.~\eqref{eq:Boltzmann_decomposed}, one finally arrives at a series of equations labeled by $l$, which depicts the coupling between different ``angular momentums'',
\begin{align}
    & 0 = \sum_{l_1,l_2} (-i)^{l_1+l_2-l} (2l_1+1)(2l_2+1)  \left(\begin{array}{ccc}
        l_1 & l_2 & l \\
        0 & 0 & 0
    \end{array}\right)^2 \nonumber\\
    & \quad\quad \times\left(
        \left( \frac{\partial \tilde{\Gamma}_{l_1}}{\partial\eta} - \frac{\partial \tilde{\Phi}}{\partial\eta} \delta_{l_1,0} \right) \Phi_\beta^{~l_2} + 
        \left( \tilde{\Gamma}_{l_1} + \tilde{\Psi} \delta_{l_1,0} \right)
        \left( \frac{({l_2}+1) s_{{l_2}+1}}{2{l_2}+1}\Phi_\beta^{~{l_2}+1} - \frac{{l_2} s_{l_2}}{2{l_2}+1}\Phi_\beta^{~{l_2}-1} \right)
    \right)~.
\end{align}
Especially, for $l=0$, the Wigner-$3$j symbol takes a simple form,
\begin{align}
    \left(\begin{array}{ccc}
        l_1 & l_2 & 0 \\
        0 & 0 & 0
    \end{array}\right) = \frac{\delta_{l_1,l_2}}{\sqrt{2l_1+1}},
\end{align}
and the corresponding projection of the Boltzmann equation is
\begin{align}
    \sum_{l=0}^{\infty} \left(
        \left( \frac{\partial \tilde{\Gamma}_{l}}{\partial\eta} - \frac{\partial \tilde{\Phi}}{\partial\eta} \delta_{l,0} \right)
        - \frac{l s_{l}}{2l+1} \left( \tilde{\Gamma}_{l-1} + \tilde{\Psi} \delta_{l-1,0} \right)
        + \frac{(l+1) s_{l+1}}{2l+1} \tilde{\Gamma}_{l+1}
    \right) \Phi_\beta^{~l} = 0~.
\end{align}
Since each $\Phi_\beta^{~l}$ is independent, the solution can be achieved by setting the factors in front of $\Phi_\beta^{~l}$ to vanish, which gives rise to the so-called hierarchy equations,
\begin{align}
    \partial_\eta \tilde{\Gamma}_0 + s_1 \tilde{\Gamma}_1 &= \partial_\eta \tilde{\Phi}, \\
    \partial_\eta \tilde{\Gamma}_1 - \frac{s_1}{3} \tilde{\Gamma}_0 + \frac{2s_2}{3}\tilde{\Gamma}_2 &= \frac{s_1}{3}\tilde{\Psi}, \\
    \partial_\eta \tilde{\Gamma}_l - \frac{ls_1}{2l+1}\tilde{\Gamma}_{l-1} + \frac{(l+1) s_{l+1}}{2l+1} \tilde{\Gamma}_{l+1} &= 0,
\end{align}
for $l=0$, $l=1$, and $l\geq 2$, respectively.

\section{Total angular momentum method}\label{app:TAM}

In this Appendix, we will introduce the total angular momentum method~\cite{Hu:1997hp}, which is the standard way of dealing with CMB anisotropies in non-flat spacetime~\cite{Hu:1997mn,Tram:2013ima,Lesgourgues:2013bra,Pitrou:2020lhu}. We apply this method to the case of CGWB anisotropies as follows to reproduce our results in the main context.

First of all, the Laplacian has a series of eigenmodes $Q^{(m)}_{i_1 i_2...i_{|m|}}$ defined as
\begin{align}
    \nabla^2 Q^{(m)}_{i_1 i_2...i_{|m|}} \equiv \gamma^{jk}D_jD_kQ^{(m)}_{i_1 i_2...i_{|m|}} = -k^2 Q^{(m)}_{i_1 i_2...i_{|m|}},
\end{align}
where $\gamma_{ij}$ represents the 3-metric of a general FLRW metric (including spatial curvature) and ``$D$'' denotes the covariant differentiation with respect to $\gamma_{ij}$. Then, we require the vector modes to be divergenceless and the tensor modes to satisfy  the transverse-traceless condition,
\begin{align}
    D^i Q_i^{(\pm 1)} &= 0,\\
    \gamma^{ij} Q_{ij}^{(\pm 2)} &= D^i Q_{ij}^{(\pm 1)} = 0.
\end{align}
In order to decompose the cosmological perturbations, we need another three kinds of auxiliary vector and tensor modes to form a complete set of basis,
\begin{align}
    Q_i^{(0)} &= -k^{-1} D_i Q^{(0)},\\
    Q_{ij}^{(0)} &= k^{-2}D_iD_jQ^{(0)} + \frac{1}{3}\gamma_{ij}Q^{(0)},\\
    Q_{ij}^{(\pm 1)} &= -(2k)^{-1}(D_jQ_{i}^{(\pm 1)} + D_iQ_{j}^{(\pm 1)}).
\end{align}
Next, we will only consider scalar perturbations, which are usually the dominant part of the CGWB anisotropies. After separating small perturbations $h_{\mu\nu}$ from the metric $g_{\mu\nu} = a^2(\gamma_{\mu\nu} + h_{\mu\nu})$, one found in Newtonian gauge
\begin{align}
    h_{00} &= -2\Psi Q^{0}, \\
    h_{ij} &= -2\Phi \gamma_{ij} Q^{0}.
\end{align}

The crucial part of the total angular momentum representation is the normal modes, which is the contraction between $Q^{(m)}$-tensors and the propagation unit vector $\hat{n}$ for the gravitons. In flat spacetime, one can express the normal modes as
\begin{align}
    G_{l}^{m} = (-i)^l \sqrt{\frac{4\pi}{2l+1}}Y_{lm}(\hat{n})e^{i\vec{k}\cdot\vec{x}},
\end{align}
with $\vec{x} = -r\hat{n}$ and $e_{3} = \hat{k}$. Then, by recognizing that the plane wave exhibits angular dependence in this coordinate system ,
\begin{align}
    e^{i\vec{k}\cdot\vec{x}} = \sum_{l} (-i)^{l}\sqrt{4\pi(2l+1)}j_{l}(kr)Y_{l0}(\hat{n}),
\end{align}
utilizing the Clebsch-Gordan relation from angular momentum theory, we finally arrive at
\begin{align}\label{glmflat}
    G_{l}^{m} = \sum_{l_1}(-i)^{l_1}\sqrt{4\pi(2l_1 +1)}j_{l_1}^{(lm)}(kr)Y_{l_1m}(\hat{n}),
\end{align}
where the specific form of $j_{l_1}^{(lm)}(kr)$ can be found in Ref.~\cite{Hu:1997hp}.

To generalize these normal modes to the curved geometry, Hu \textit{et al.}~\cite{Hu:1997mn} realise that one can construct the normal modes with the same structure in non-flat spacetime,
\begin{align}
    G_{l}^{m} = (-i)^l \sqrt{\frac{4\pi}{2l+1}}Y_{lm}(\hat{n})e^{i\delta (\vec{x}, \vec{k})},
\end{align}
where $\delta (\vec{x}, \vec{k})$ is some scalar function, and can in principle be calculated with the recursion formula of $Q^{(m)}$,
\begin{align}
    \hat{n}^{i}D_i(G_{l}^{m}) = \frac{q}{2l+1}[\kappa_{l}^{m}(G_{l-1}^{m}) - \kappa_{l+1}^{m}(G_{l+1}^{m})], 
\end{align}
with $q=\sqrt{k^2+(|m|+1)K}$ and the coupling coefficient
\begin{align}
    \kappa_{l}^{m} = \sqrt{(l^2-m^2)\left(1-\frac{l^2}{q^2}K \right)}.
\end{align}
Then one can obtain a similar relation to~(\ref{glmflat})
\begin{align}
    G_{l}^{m} = \sum_{l_1}(-i)^{l_1}\sqrt{4\pi(2l_1+1)}\phi_{l_1}^{(lm)}Y_{l_1m}(\hat{n}),
\end{align}
where the specific form of $\phi_{l_1}^{(lm)}$ can be found in Ref.~\cite{Hu:1997mn}.

Now one can directly calculate the Boltzmann equation,
\begin{align}\label{boltzeq}
    \frac{\md \Gamma}{\md \eta} = \frac{\partial \Gamma}{\partial \eta} + \hat{n}^{i}D_{i}\Gamma = \mathcal{C}[\Gamma] +\mathcal{T}[\Gamma] + \mathcal{G}[h_{\mu\nu}] ,
\end{align}
where $\mathcal{C}[\Gamma]$ is the collision term and it will be ignored in the calculation of CGWB anisotropies, while $\mathcal{T}[\Gamma]$ is the emission term acting as an initial condition, and the gravitational redshifts term $\mathcal{G}[h_{\mu\nu}]$ is expressed as
\begin{align}
    \mathcal{G}[h_{\mu\nu}] = -\frac{1}{2}\hat{n}^{i}\hat{n}^{j}h_{ij}^{\prime} - \hat{n}^{i}h_{0i}^{\prime} + \frac{1}{2}\hat{n}^{i}D_i h_{00}.
\end{align}

Finally, after expanding $\Gamma(\eta, \vec{x}, \hat{n}, p)$ in normal modes $G_{l}^{m}$,
\begin{align}\label{gammak}
    \Gamma(\eta, \vec{x}, \hat{n}, p) = \int\frac{\md^3q}{(2\pi)^3}\sum_{l}\tilde{\Gamma}_{l}(\eta, \vec{k}, \hat{n}, p)G_{l}^{m},
\end{align}
and plugging~(\ref{gammak}) into~(\ref{boltzeq}), one arrive at the hierarchy equations of $\tilde{\Gamma}$,
\begin{align}
    \tilde{\Gamma}_{l} = q\left[\frac{\kappa_{l}^{m}}{(2l-1)}\tilde{\Gamma}_{l-1} -  \frac{\kappa_{l+1}^{m}}{(2l+3)}\tilde{\Gamma}_{l+1}\right] + \tilde{S}_{l},
\end{align}
with the scalar sources $\tilde{S}_{l}$ expressed as
\begin{align}
    \tilde{S}_{0} = \tilde{\Gamma}_{0}\delta(\eta-\eta_{\mathrm{in}})+\tilde{\Phi}^{\prime}, ~~~~~\tilde{S}_{1} = k\tilde{\Psi},
\end{align}
where $\tilde{\Gamma}_{0}(\eta_{\mathrm{in}}, \vec{k}, \hat{n}, p)$ is the integration constant and it is also the initial condition of $\tilde{\Gamma}_{l}(\eta, \vec{k}, \hat{n}, p)$~\footnote{The higher order terms are neglected for they are subdominant.}. Hence, the integral solution follows,
\begin{align}
    \tilde{\Gamma}_{l}(\eta_{0}, \vec{k}, \hat{n}, p) = \left(\int_{\eta_{\mathrm{in}}}^{\eta_0} \md\eta \sum_{j}\tilde{S}_{j}\phi_{l}^{(j0)}\right),
\end{align}
where $\Phi^{~l}_{\beta}(\chi)$ is the hyper-spherical Bessel function. After integrating by parts, one can obtain the well-known form,
\begin{align}
    \tilde{\Gamma}_{l}(\eta, \vec{k}, \hat{n}, p) = \int_{\eta_{\mathrm{in}}}^{\eta_0} \md\eta\left[(\tilde{\Phi}^{\prime}+\tilde{\Psi}^{\prime})+(\tilde{\Psi}+\tilde{\Gamma}_{0})\delta(\eta-\eta_{\mathrm{in}})\right]\Phi^{~l}_{\beta}(\chi),
\end{align}
where $\chi = \sqrt{|K|}(\eta_{0} - \eta)$. This formula is exactly the same as~\eqref{eq:Gammal_SW_and_ISW}.

\bibliographystyle{JHEP}
\bibliography{ref}

\end{document}